\documentclass{aa}

\usepackage{graphicx}
\usepackage{txfonts}
\usepackage{xcolor}
\usepackage{ulem}
\usepackage{url}
\usepackage{longtable}
\usepackage{multirow}
\usepackage{siunitx}
\usepackage{booktabs}
\usepackage{amsmath}
\usepackage{adjustbox}
\usepackage{orcidlink}
\usepackage{lscape}
\usepackage{tablefootnote}
\usepackage{listings}
\makeatletter
\renewcommand*\aa@pageof{, page \thepage{} of \pageref*{LastPage}}
\begin{document}

   \title{The Brighter-Fatter Effect in the JWST MIRI Si:As IBC detectors}
   \subtitle{I. Observations, impact on science, and modelling}

   \author{Ioannis~Argyriou\inst{1}\orcidlink{0000-0003-2820-1077}
          \and
          Craig~Lage\inst{2}\orcidlink{0000-0002-9601-345X}
          \and
          George~H.~Rieke\inst{3}\orcidlink{0000-0003-2303-6519}
          \and
          Danny~Gasman\inst{1}\orcidlink{0000-0002-1257-7742}
          \and
          Jeroen~Bouwman\inst{4}\orcidlink{0000-0003-4757-2500}
          \and
          Jane~Morrison\inst{3}\orcidlink{0000-0002-9288-9235}
          \and 
          Mattia~Libralato\inst{5}\orcidlink{0000-0001-9673-7397}
          \and 
          Daniel~Dicken\inst{6}\orcidlink{0000-0003-0589-5969}
          \and 
          Bernhard~R.~Brandl\inst{7,8}\orcidlink{0000-0001-9737-169X}
          \and
          Javier~Álvarez-Márquez\inst{9}\orcidlink{0000-0002-7093-1877}
          \and
          Alvaro~Labiano\inst{9,10}\orcidlink{0000-0002-0690-8824}
          \and
          Michael~Regan\inst{11}\orcidlink{0000-0001-9367-0705}
          \and
          Michael~E.~Ressler\inst{12}\orcidlink{0000-0001-5644-8830}
         }

   \institute{Institute of Astronomy, KU Leuven,
              Celestijnenlaan 200D, 3001 Leuven, Belgium
              \and
              Department of Physics, University of California-Davis, 1 Shields Ave. Davis, Ca., U.S.A.
              \and
              Steward Observatory and the Department of Astronomy, The University of Arizona, 933 N Cherry Ave, Tucson, AZ, 85750, USA
              \and
              Max Planck Institute for Astronomy, K\"{o}nigstuhl 17, D-69117 Heidelberg, Germany
              \and
              AURA for the European Space Agency (ESA), Space Telescope Science Institute, 3700 San Martin Drive, Baltimore, MD 21218, USA
              \and
              UK Astronomy Technology Centre, Royal Observatory Edinburgh, Blackford Hill, Edinburgh EH9 3HJ, UK
              \and
              Leiden Observatory, Leiden University, PO Box 9513, 2300 RA Leiden, The Netherlands
              \and
              Faculty of Aerospace Engineering, Delft University of Technology, Kluyverweg 1, 2629 HS Delft, The Netherlands
              \and 
              Telespazio UK for the European Space Agency, ESAC, Camino Bajo del Castillo s/n, 28692 Villanueva de la Ca\~nada, Spain \label{tpz}
              \and
              Centro de Astrobiolog\'{\i}a (CAB), CSIC-INTA, Ctra. de Ajalvir km 4, Torrej\'on de Ardoz, E-28850, Madrid, Spain \label{cab}
              \and
              Space Telescope Science Institute, 3700 San Martin Drive, Baltimore, MD, 21218, USA
              \and
              Jet Propulsion Laboratory, California Institute of Technology, 4800 Oak Grove Dr., Pasadena, CA, 91109, USA\\
              \email{ioannis.argyriou@kuleuven.be}
         }

   \date{Received \today}
   
   \titlerunning{JWST MIRI Brighter Fatter Effect}

   \authorrunning{I. Argyriou, C. Lage, G. H. Rieke et al.}


  \abstract
    {The Mid-Infrared Instrument (MIRI) on board the James Webb Space Telescope (\textit{JWST}) uses three Si:As impurity band conduction (IBC) detector arrays. The output voltage level of each MIRI detector pixel is digitally recorded by sampling-up-the-ramp. For uniform or low-contrast illumination, the pixel ramps become non-linear in a predictable way, but in areas of high contrast, the non-linearity curve becomes much more complex. The origin of the effect is poorly understood and currently not calibrated.}
   {We provide observational evidence of the Brighter-Fatter Effect (BFE) in MIRI conventional and high-contrast coronographic imaging, low-resolution spectroscopy, and medium-resolution spectroscopy data and investigate the physical mechanism that gives rise to the effect on the MIRI detector pixel raw voltage integration ramps.}
   {We use public data from the \textit{JWST}/MIRI commissioning and Cycle 1 phase. We also develop a numerical electrostatic model of the MIRI detectors using a modified version of the public Poisson\_CCD code.}
   {We find that the physical mechanism behind the BFE manifesting in MIRI data is fundamentally different to that of CCDs and photodiode arrays such as the Hawaii-XRG (HXRG) near-infrared detectors used by the NIRISS, NIRCam, and NIRSpec instruments on board \textit{JWST}. Observationally, the BFE makes the \textit{JWST} MIRI data yield 10~--~25~\% larger point sources and spectral line profiles as a function of the relative level of debiasing of neighboring detector pixels. This broadening impacts the MIRI absolute flux calibration, time-series observations of faint companions, and PSF modeling and subtraction. We also find that the intra-pixel 2D profile of the shrinking Si:As IBC detector depletion region directly impacts the accuracy of the pixel ramp non-linearity calibration model.}
   {}



  \maketitle
%

\section{Introduction}
\label{sec:introduction}

The Mid-Infrared Instrument MIRI \citep{wright23}, on board the \textit{James Webb Space Telescope (JWST)}, has four operational modes, (1) imaging, (2) low-resolution spectroscopy, (3) high-contrast (coronagraphic) imaging, and (4) medium-resolution spectroscopy \citep{miri_pasp_2, miri_pasp_3,coronograph_perf,miri_pasp_5,Wells_2015}. The in-flight performance of MIRI and its four operational modes (including time-series observations) is presented in \citet{wright23,dicken23,kendrew23,bouwman23,coronograph_perf,argyriou23}. MIRI uses three Si:As impurity band conduction (IBC) detector arrays, one for conventional imaging, high-contrast imaging and low-resolution spectroscopy, and two for medium-resolution spectroscopy \citep{miri_pasp_7}. Si:As IBC devices have extensive space flight heritage, having been used for example in all three instruments of the Spitzer space telescope, namely the Infrared Array Camera (IRAC), the Infrared Spectrograph (IRS), and the Multiband Imaging Photometer (MIPS). The high quantum efficiency of the Si:As IBC devices, in combination with the extensive wavelength range covered (5-28~$\mu m$) gives these devices a unique advantage \citep{love2005,Gaspar2021}. Other detectors, such as the Teledyne Imaging Sensors' LWIR HgCdTe detector have a relatively higher quantum efficiency, however, they cover a shorter wavelength range, namely out to 13~$\mu m$\footnote{The Teledyne Imaging Sensors' LWIR HgCdTe detectors were not available at the time of the detector selection for MIRI.} \citep{Dorn2018}.

The Brighter-Fatter Effect (BFE) is a well-known detector effect that blurs the intrinsic spatial and spectral information of a source. The impact of the effect on Charge-Coupled Devices (CCDs) has been studied in detail observationally, experimentally in a laboratory environment, and by electrostatic modeling \citep{antilogus14,guyonnet2015,Lage2017,Coulton2018}. More recently the BFE has been studied in similar detail for the Hawaii-XRG (HXRG) HgCdTe photodiode arrays operating in the near-infrared regime \citep{plazas17,Plazas2018,hirata2020}. The NIRISS, NIRCam, and NIRSpec instruments on board JWST all use H2RG detectors manufactured by Teledyne Imaging Systems \citep{doyon23,marcia23,boker23}. There is no literature on the presence in - or impact of the BFE on - Si:As IBC devices. Indeed there is a fundamental question of whether BFE can manifest in Si:As IBC devices at all. In this first paper on the MIRI BFE we report on the observation of the BFE in MIRI data in all four operational modes of MIRI, its impact on MIRI science, as well as present a model that demonstrates how the BFE arises inside the Si:As IBC devices. In the second paper in the series (Gasman et al., in prep.) we present a self-calibration algorithm to correct the BFE in MIRI data.

In Sect.~\ref{sec:detectors} of this paper we describe how the MIRI detector arrays work, how they produce the raw output signal used to estimate the flux from astronomical sources, and how the detector pixel non-linear behavior to non-uniform illumination affects the raw output signal. In Sect.~\ref{sec:observations} we show and describe how the MIRI BFE impacts the scientific interpretation of MIRI imaging, low-resolution (time-series) spectroscopy, medium-resolution spectroscopy and high contrast (coronographic) imaging. In Sect.~\ref{sec:modeling} we present a realistic electrostatic model of the MIRI detectors, that illustrates the physical mechanism behind the effect. For this we use the public \texttt{Poisson\_CCD} numerical code based on \citet{Lage2021}. We also present the results of the electrostatic simulations, limiting the simulations to the imaging case. In Sect.~\ref{sec:discussion} the implications of the presented results are discussed in the context of how the BFE manifests in CCD and HXRG detectors and the difference in the physical mechanism of the BFE between the different detectors. The conclusions of this study are formulated in Sect.~\ref{sec:conclusions}.

\begin{figure*}[t]
\centering
\includegraphics[width=0.97\textwidth]{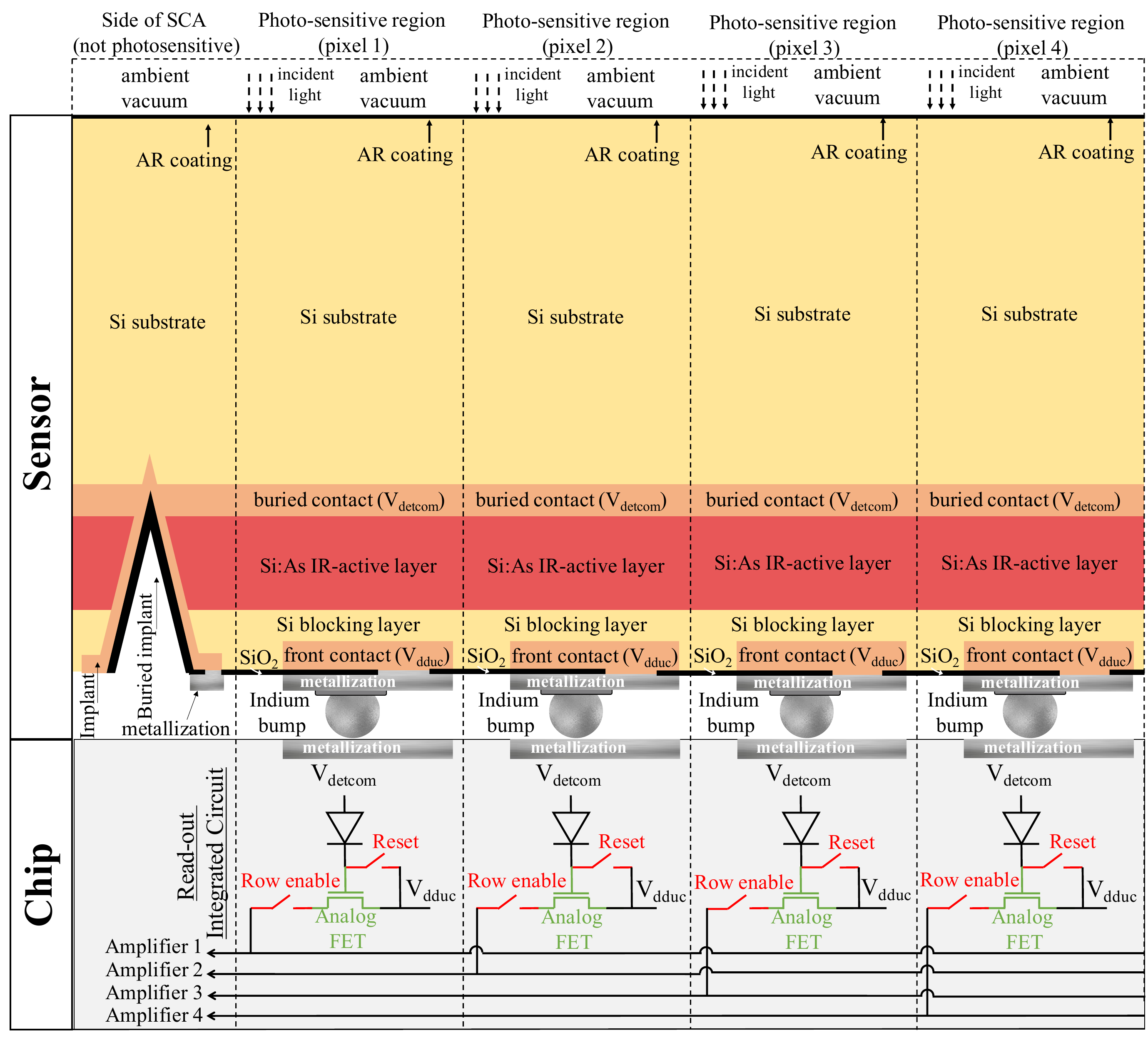}
\caption[example]{Representation of MIRI detector architecture based on \cite{petroff_stapelbroek_patent,love2005,miri_pasp_7,Gaspar2021}]. The dimensions are not to scale.}
\label{fig:detector_layout}
\end{figure*}

\begin{figure}[ht!]
\centering
\includegraphics[width=0.49\textwidth]{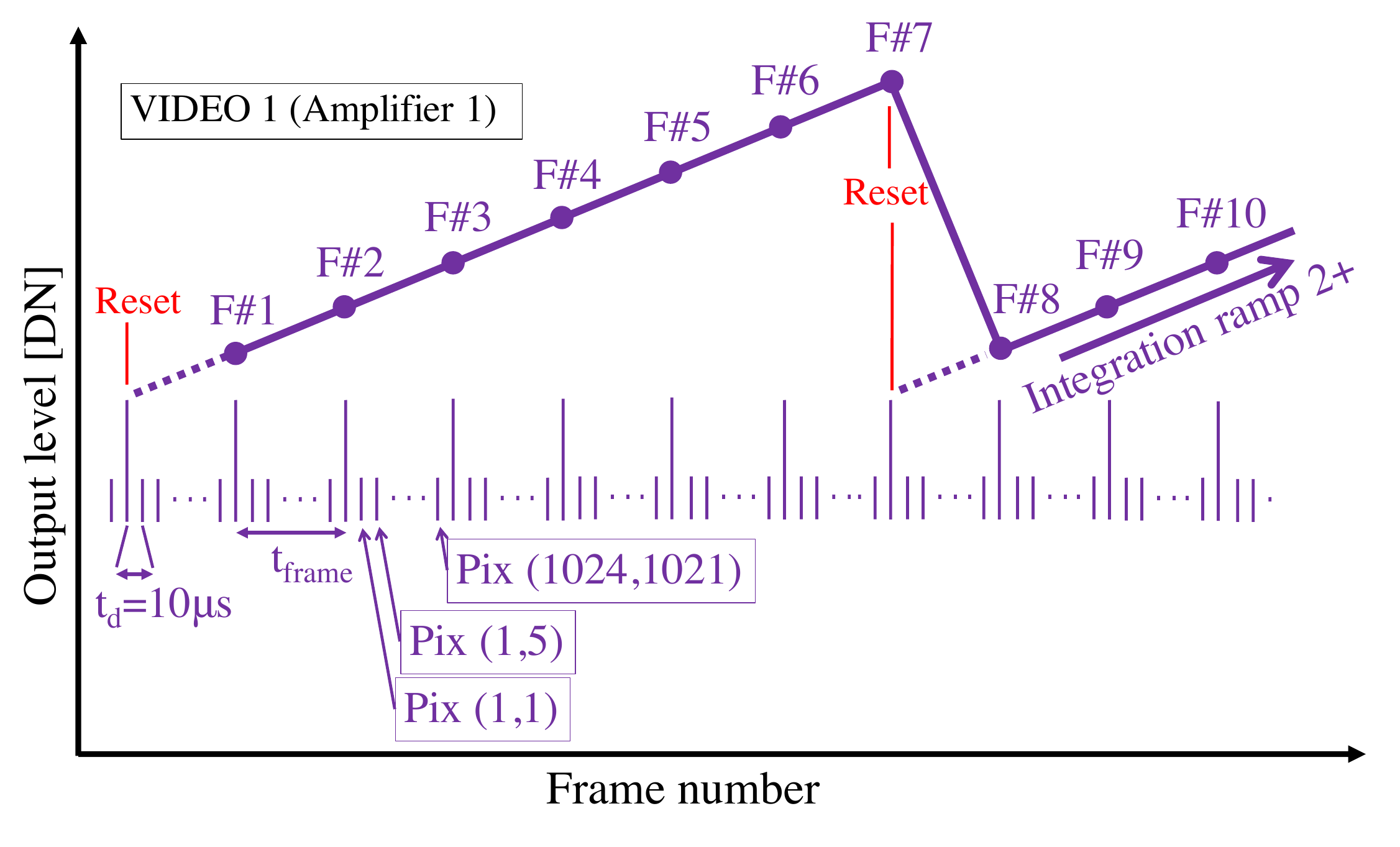}
\caption{Conceptual representation of a MIRI detector pixel integration ramp. Individual frames are designated by F\#.}
\label{fig:miri_integration_ramp}
\end{figure}

\section{The \textit{JWST} MIRI detectors and the pixel non-linear response to non-uniform illumination}
\label{sec:detectors}

\subsection{The detector architecture}

To understand the MIRI BFE, an in-depth understanding of the Si:As IBC detector architecture and read-out integration circuit (ROIC) is required. We describe these first, starting from a representation of the architecture of the MIRI detector arrays, shown in Fig.~\ref{fig:detector_layout}. Si:As IBC detectors are grown on a silicon substrate. Photons pass through the anti-reflection coating on the detector back side, into the substrate, and then through the ion-implanted buried contact into the infrared-active layer (detection layer). The infrared-active layer is doped with arsenic to absorb the incoming photons, which elevate the photo-excited electrons (henceforth photoelectrons) from the impurity band into the conduction band. Assuming a low level of minority acceptor impurities in this layer, an electric field can be maintained across it that causes the photoelectrons to migrate to the frontside of the detector. A thin, high purity layer (blocking layer) is grown over the frontside of the infrared-active layer. When operated at sufficiently low temperatures, thermally-generated free charge carriers cannot penetrate the blocking layer because it lacks an impurity band and the carriers have insufficient energy to be lifted into the conduction band (a band diagram for front-illuminated Si:As IBC detectors can be found in \citet{rieke2003}, adapted as Fig. 2.2 in \citet{phdthesisYannis}). The photo-generated free charge carriers in the conduction band can, however, traverse the blocking layer, to be collected at the detector frontside contact. The electric field that drives this process is maintained across the infrared-active layer between the frontside contact on the readout side, and the buried contact, connected via a V-etch buried implant on one side of the detector. The photoelectrons are transferred through an ion-implanted transition region to a metalized output pad. The indium bump conveys the signal to a matching input pad on a readout amplifier. The MIRI detector operational temperature of 6.4~K keeps the detector dark current very low (<~0.1~e$^-$/s) \citep{miri_pasp_7}.

The MIRI detector bias voltage $V_{bias}$ is defined by Eq.~\ref{eq:bias_voltage} \citep{miri_pasp_8}.

\begin{equation}\label{eq:bias_voltage}
    V_{bias} = V_{dduc} - V_{detcom} + 0.2V
,\end{equation}

\noindent where $V_{dduc}$ is the voltage of the frontside contact, and $V_{detcom}$ is the voltage of the buried contact. The bias voltage is set whenever the reset switch is closed. After the reset switch is opened, the pixels start integrating signal and the bias voltage drops. The voltage from the detector is buffered by an analog Field Effect Transistor (FET) within the pixel unit cell.

The signal is passed to the output by operating the row and column switches (a single row is shown in Fig.~\ref{fig:detector_layout}). Shown at the bottom of Fig.~\ref{fig:detector_layout} are the five MIRI detector video lines, each with its own read-out amplifier. Amplifiers 1 to 4 (video lines 1 to 4) address the first to fourth pixel columns from left to right, respectively. In the case of permanent loss of one of these four video lines, this would result in the loss of an alternating fourth column of array imagery.

\subsection{The MIRI detector pixel read-out}\label{subsec:pixel_readout}

The MIRI focal plane module (FPM) focal plane electronics (FPE) measure signal in the form of integration ramps. These ramps are the raw data that will be discussed throughout this paper. MIRI pixels are read non-destructively (charge is read but not reset). An illustration of an ideal pixel integration ramp is shown in Fig.~\ref{fig:miri_integration_ramp}. On the x-axis is the image frame number, and on the y-axis is the output level of a video line. The read-out electronics measure the voltage difference as Digital Numbers (DN). The proportionality between the DN and the voltage difference is defined by the system gain.

Whenever the detectors are idle (non-exposing) the pixels are constantly resetting at the frame rate. In this case the amplifier integrating node voltage remains close to the bias voltage. An exposure starts when the reset switch is opened, as shown in Fig.~\ref{fig:detector_layout}. At that point, with the reset switch kept open, the amplifier integration node voltage changes based on the photoelectric current. The current depends on the number of photons detected. The time difference between two pixels being read out is set by the FPE master 100~kHz clock ($t_d=10$~$\mu s$) \citep{miri_pasp_8}. After the first pixel is read out (pixel (1,1) in Fig.~\ref{fig:miri_integration_ramp}), all other pixels on the detector have to be read out before the first pixel can be read out again. The five amplifiers are read out at the same time. Specifically, taking row 1 as an example, pixels (1,5,9,...,1021) addressed by Amplifier~1, are read at the same time as pixels (2,6,10,...,1022) addressed by Amplifier~2, and so on. Every time the detector is read out in its entirety, a detector frame is stored (equivalent to one frame number in Fig.~\ref{fig:miri_integration_ramp}). At the end of an integration, the reset switch is closed and the node voltage returns to the bias voltage. MIRI observers can take multiple integrations in a single exposure, and in that case, after a virtually instantaneous reset, the pixel starts integrating signal again. The default operation of the MIRI FPE in flight is to perform two reset frames in succession to mitigate reset switch charge decay effects, i.e., once to remove the integrated signal and again to set a solid zero point \citep{morrison23}.

A 16 bit ADC (Analog to Digital Converter) in the readout electronics converts the voltage in $2^{16}$ different values, i.e. 65536 different values going from 0~DN up to 65535~DN. The MIRI detector output is approximately +0.5~Volt for zero signal and -1.0~Volt for saturation. Ideally, the detector +0.5~V would be read as 0~DN, and -1.0~V would be read as 65535~DN, however, this is not the case, a margin is necessary, related to the spread in the readout chain and the detector characteristics. As such, the detector range is not mapped to the full analog-to-digital converter range. \cite{miri_pasp_8}, estimated a net system gain of the MIRI detectors of 38300~DN/V, so +0.5~V to -1.0~V changes 57450~DN. \cite{miri_pasp_7} report that "the detector is a simple plane-parallel 1.1~fF capacitor, the indium bump bonds may contribute an additional $\sim$4~fF to the node capacitance \citep{moore2005}, and the capacitance at the input to the [pixel] unit cell buffer amplifier is 28.5~fF \citep{mcmurtry2005}, thus the nominal capacitance at the integrating node is 33.6~fF". Using these capacitance values and the net system gain, \cite{miri_pasp_7} derived a pixel gain of 5.5~e$^-$/DN. This value was uncertain due to a number of effects, particularly interpixel capacitance. Experience in flight has indicated a smaller pixel gain value, but this is still work in progress. In this paper we use the pixel gain value of 5.5~e$^-$/DN. 

\subsection{The non-linearity of the MIRI detector pixel voltage integration ramps}\label{subsec:miri_nonlinearity}

\begin{figure}[t]
\centering
\includegraphics[width=0.48\textwidth]{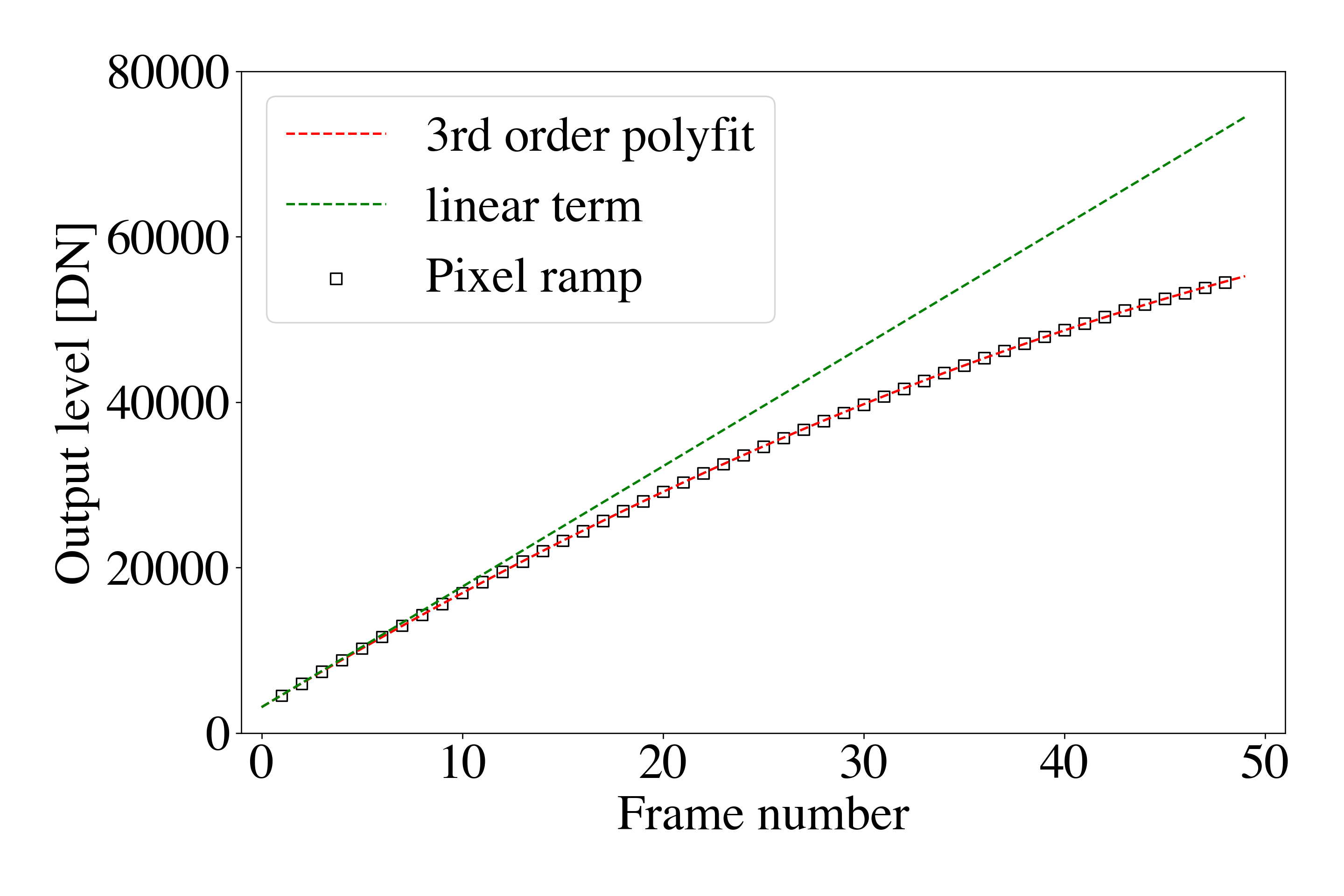}
\caption[Non-linearity in MIRI pixel ramp.]{Non-linearity in MIRI pixel ramp (black squares). A third order polynomial is fitted to the ramp (red dashed line). The dashed green line shows the linear term of the third order fit.}
\label{fig:non_linearity}
\end{figure}

Similar to most other detectors used in astronomical observing applications, the MIRI detector pixel integration ramps suffer from a non-linear response as a function of the DN output level. An example of a MIRI pixel ramp is shown in Fig.~\ref{fig:non_linearity}. Instead of the output level increasing linearly with frame number (time) the slope of the ramp decreases. Traditionally a correction for this non-linearity is derived by (1) fitting a polynomial to the ramp, (2) using the linear term of the fit as a proxy for what the ramp would look like if not impacted by the cause behind the non-linearity, (3) defining a correction as the ratio of the linear term and the polynomial fit, and that as a function of the DN level at each frame time \citep{morrison23}.

On a system level, for the MIRI Si:As IBC detectors, non-linearity is caused by a reduction in detector responsivity with increasing charge accumulation at the amplifier integrating node capacitance \citep{miri_pasp_7}. A voltage $V_{dduc}=-2.0$~V is applied at the frontside contact and a voltage $V_{detcom}=-4.0$~V is applied at the detector buried contact. An additional 0.2~V is due to clock feedthrough (see \citet{miri_pasp_8}) for a total detector bias voltage $V_{bias}=2.2$~V, as formulated in Eq.~\ref{eq:bias_voltage}. Under these conditions the region between the frontside contact and the buried contact becomes fully depleted of free charge carriers. Photoelectrons are produced throughout the depletion region (infrared-active layer) and the electric field causes them to migrate towards the pixels. The MIRI detector bias voltage of 2.2~V was tuned for the width of the depletion region to cover the entire active layer without causing avalanche gain \citep{miri_pasp_7}. However, charge accumulation at the amplifier integrating node capacitance reduces the net bias voltage on the detector. As a result, at higher DN levels the net bias voltage decreases, the width of the depletion region shrinks below the active layer width, and a smaller fraction of the produced photoelectrons are detected -- recombination can prevent photoelectrons from reaching the depleted region -- resulting in a reduction in detector responsivity as a function of DN.  

The 1D equation for the width of the MIRI detector depletion region is mathematically described by Eq.~\ref{eq:width_of_depletion_region} \citep{pierret1996,miri_pasp_7}.

\begin{figure}[t]
\centering
\includegraphics[width=0.48\textwidth]{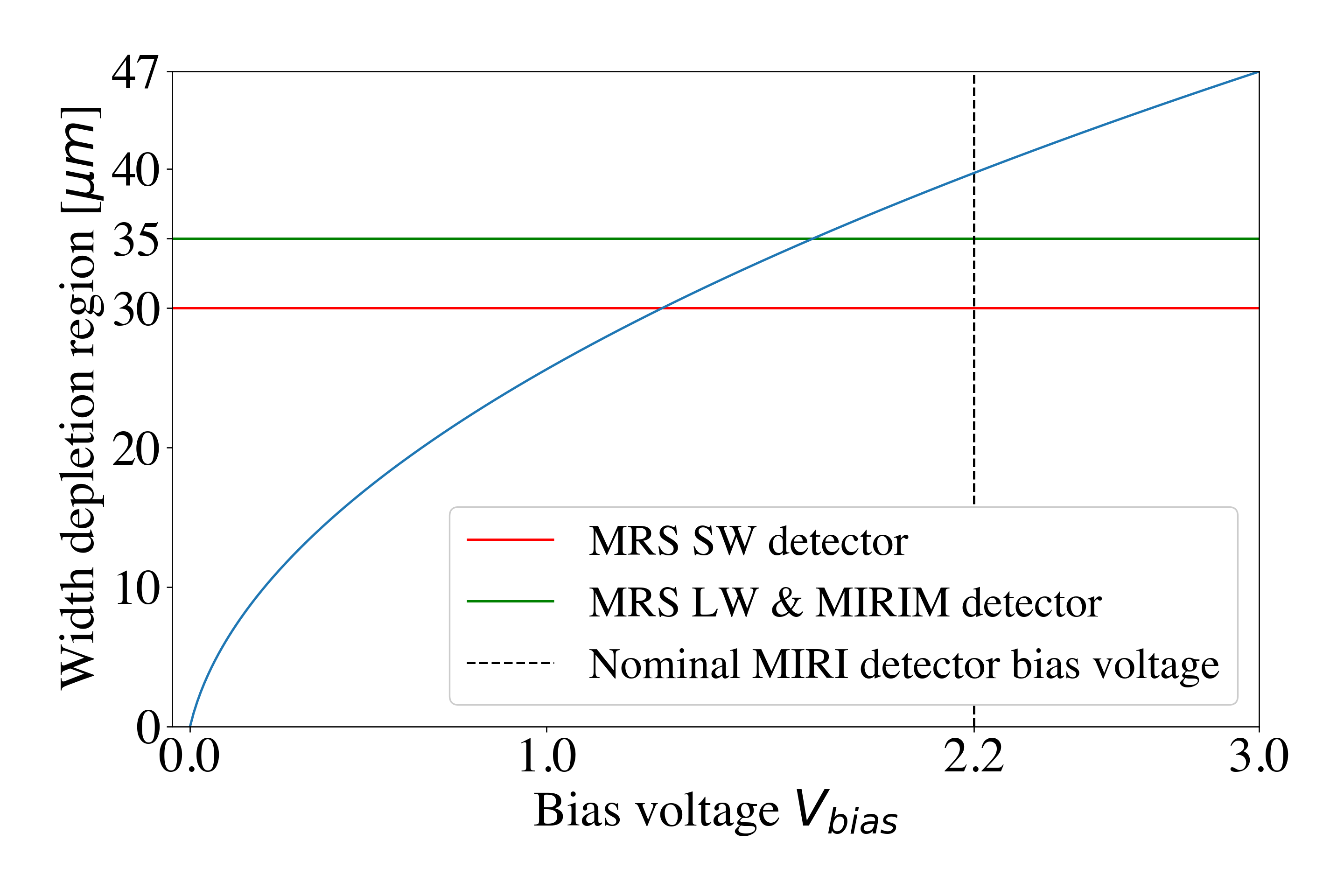}
\caption{Width of MIRI detector depletion region as a function of bias voltage (blue curve). The horizontal red and green lines represent the width of the depletion region for the MRS SW detector and the MRS LW and MIRI imager detectors respectively. The nominal MIRI detector bias voltage is 2.2~V for all three MIRI detectors \citep{miri_pasp_8}.}
\label{fig:width_of_depletion_region}
\end{figure}

\begin{equation}\label{eq:width_of_depletion_region}
w=\left[\frac{2 \kappa_{0} \varepsilon_{0}}{q N_{A}}\left|V_{bias}\right|+t_{B}^{2}\right]^{1 / 2}-t_{B}
,\end{equation}

\begin{figure*}[ht!]
\centering
\includegraphics[width=0.98\textwidth]{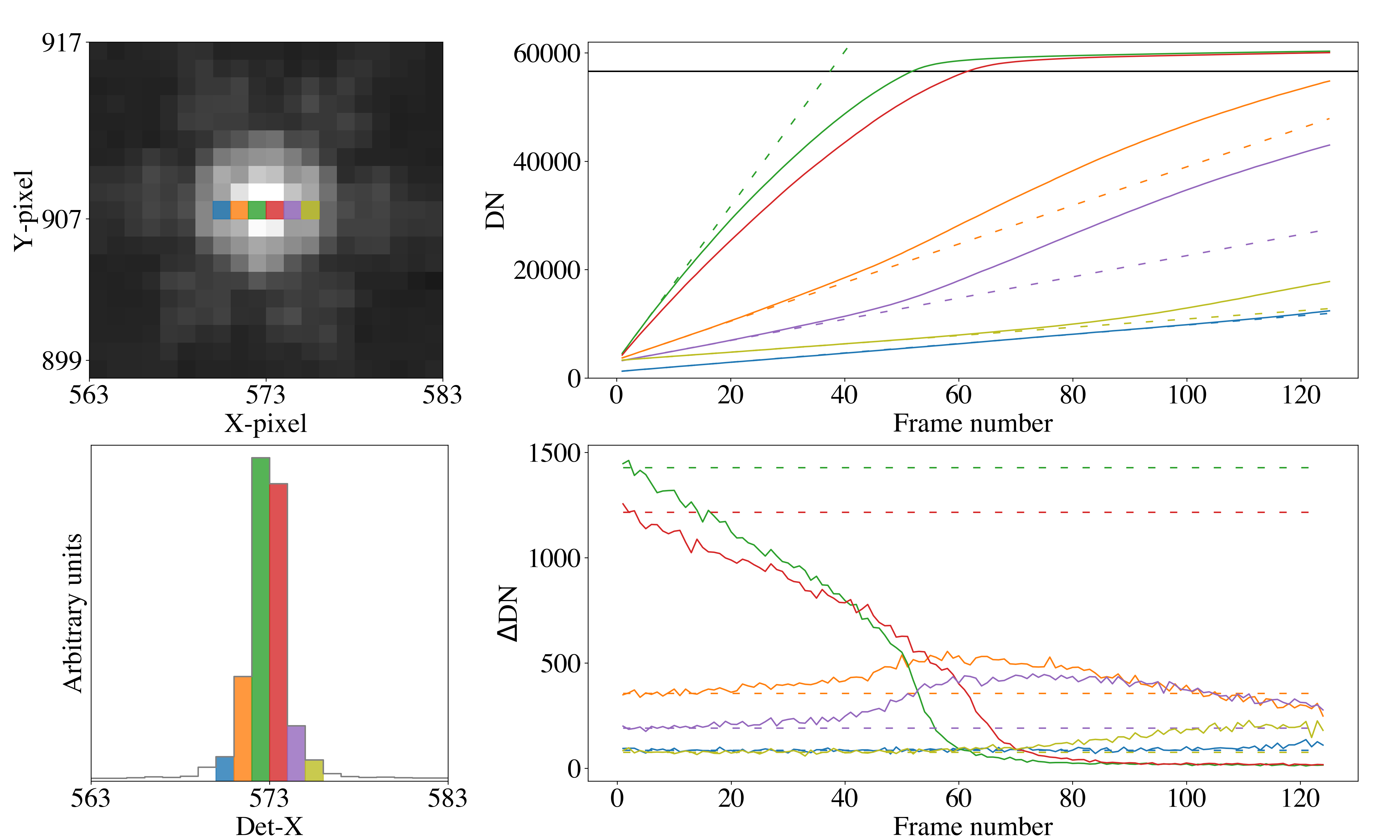}
\caption{MIRI F560W imager data of an unresolved point source observed during the \textit{JWST} commissioning phase. Top left: Reduced MIRI imaging data plotted in log scale. Six pixels are selected in detector row 907 across the horizontal width of the PSF (colored boxes). Bottom left: Signal distribution across detector row 907. The same color is used for the selected pixels in the top left panel. Top right: Raw pixel voltage integration ramps of the pixels selected in the top left panel (solid colored lines). The loosely dashed colored lines are linear regressions on the first 5 frames of the selected ramps. The blue ramp has been shifted down by 2000~DN for clarity. The horizontal black line at 56.600~DN shows the saturation value for the MIRI imager pixels. Bottom right: Subsequent frame differences of ramps shown in the top right panel ($\Delta DN = DN_{i+1} - DN_{i}$). The loosely dashed colored lines are the slopes determined from the linear regressions on the first 5 frames of the selected ramps.}
\label{fig:bfe_signature_imager}
\end{figure*}

\noindent here $w$ is the width of the depletion region, $\kappa_{0}=11.68$ is the dielectric constant of silicon, $\varepsilon_{0}=8.854 . 10^{-12}$~$\frac{C}{V*m}$ is the permittivity of free space, $q=1.6 . 10^{-19}$~C is the elementary charge, $N_{A}=1.5 . 10^{12}$~cm$^{-3}$ is the density of the minority (acceptor) impurity concentration in the MIRI detector active layer, $V_{bias}=2.2$~V is the bias voltage, and $t_{B}=4$~$\mu m$ is the width of the detector blocking layer \citep{love2005,miri_pasp_7}. We use these values to plot the width of the depletion region $w$ as a function of bias voltage $V_{bias}$ in Fig.~\ref{fig:width_of_depletion_region}. Based on the provided values we find that for a bias voltage $V_{bias}$ = 2.2~V the depletion region covers the full MRS short wavelength (SW), long wavelength (LW) detectors, and imaging (MIRIM) detector.

For a DN value of 50000~DN, if we use the net system gain value of 38300~DN/V of \cite{miri_pasp_7}, we estimate a debiasing of $\sim$1.3~V. Tracing this back to Fig.~\ref{fig:width_of_depletion_region}, at that level of debiasing (x-axis value of 0.9~V) the width of the depletion region is much less than the active layer width of both the MRS SW detector, LW detector, and MIRIM detector. Photoelectrons in the undepleted region can still be collected if they diffuse to the depleted region, but at reduced efficiency. This is the understood cause of the non-linearity at higher DN shown in Fig.~\ref{fig:non_linearity}. Calibrating the non-linearity effect due to the detector debiasing is equivalent to predicting what the output DN level would be if the active layer was always fully depleted.

The non-linearity is wavelength-dependent due to the absorption coefficient of the Si:As infrared-active layer being wavelength-dependent. Up to 15~$\mu m$ the absorption coefficient is well described by Eq.~\ref{eq:absorption_coefficient} \citep{woods2011,miri_pasp_7}.

\begin{equation}
    \alpha(\lambda) = 102*(\lambda/7\mu m)^2 \, cm^{-1}
\label{eq:absorption_coefficient}
\end{equation}

The quadratic dependence in Eq.~\ref{eq:absorption_coefficient} implies that longer-wavelength photons have a higher probability of being absorbed closer to the buried contact. That in combination with the decrease in the width of the depletion region towards the detector frontside (Eq.~\ref{eq:width_of_depletion_region}) means that the reduction in response is greater at longer wavelengths. This is quantified and shown using imager data in different wavelength filters in \citet{morrison23}.

\subsection{The issue with defining a global MIRI non-linearity solution for all illumination profiles}

For the MIRI imager, for a single wavelength filter, the \textit{JWST} MIRI data calibration pipeline uses a non-linearity solution that is common for all the pixels on the detector. Assuming a perfectly uniform detector illumination and a perfectly uniform response for all pixels, this would imply a single global correction, derived as per the process described in Sect.~\ref{subsec:miri_nonlinearity} and illustrated in Fig.~\ref{fig:non_linearity}. In this simple case the uncertainties on the non-linearity solution would be linked to read-out noise, photon-noise, dark current noise, and the reset effects impacting the start of MIRI ramps \citep{morrison23}. Although one would expect the global "classical" non-linearity solution to hold for all types of illumination -- and indeed it does for low-contrast extended and low-contrast semi-extended sources -- it fails in regions of high contrast, for example for point sources. This is notably of concern at the shorter MIRI wavelengths (5-15 $\mu m$) because at longer wavelengths (above 15~$\mu m$) the JWST thermal background increasingly reduces the signal contrast between the science target and the surrounding background.

In Fig.~\ref{fig:bfe_signature_imager}, in the top left panel, we show a small section of the MIRI imager field of view where a bright unresolved (point) source was observed in flight during a multi-instrument multi-field \textit{JWST} in-focus test (commissioning program ID (PID) \href{https://www.stsci.edu/jwst/science-execution/program-information.html?id=1464}{1464}). Here, the MIRI imager F560W wavelength filter (centered on 5.6~$\mu m$) was used and the \textit{JWST} telescope Point Spread Function (PSF) in the F560W filter can be discerned. We select six pixels that cross the core of the PSF in the horizontal direction, and plot the signal distribution (arbitrary units) in the bottom left panel. Each pixel measures a raw voltage integration ramp as shown in the top right panel of Fig.~\ref{fig:bfe_signature_imager} as solid colored curves\footnote{The green-colored ramp is the same ramp we showed in Fig.~\ref{fig:non_linearity}; now we show the full extent of the ramp going well past saturation.}. The loosely dashed colored lines are linear fits to the first 5 frames of the raw ramps, where the debiasing effect on the non-linearity is smaller. The black horizontal line shows the saturation level of the MIRIM detector. Specifically, for the MIRI imager, coronographs, and low-resolution spectrometer mode, any frame measuring a signal level higher than 56.660~DN is flagged as saturated and is not used in the ramp slope fitting that determines the pixel-incident flux. For the two detectors used by MIRI medium-resolution spectrometer, the saturation value is 55.000~DN.

The green and red-colored pixels in the bottom right panel of Fig.~\ref{fig:bfe_signature_imager} record a similar output level and display a non-linear behavior in the top right panel that is well understood in the context of a uniform detector illumination. However, the ramp of the orange-colored pixel located one position to the left of the green pixel on the detector shows a signature that deviates strongly from the classical understanding of the MIRI ramp non-linearity. Instead of a decrease in slope, the slope increases. Clearly, the solution that linearizes the green-colored ramp cannot linearize the orange-colored ramp. Quite the contrary, correcting the orange-colored ramp with the same solution as the green ramp will result in the orange-colored ramp curving even more significantly upwards. Given that the \textit{JWST} calibration pipeline determines the flux of a source by performing a linear regression on the non-linearity corrected frames below the saturation limit, it is important to correctly linearize the ramps first and foremost.

In the bottom right panel of Fig.~\ref{fig:bfe_signature_imager} we show with the solid curves the subsequent frame differences of the ramps of the selected pixels. The loosely dashed lines illustrate the estimated slopes based on the output level in the first five frames of the raw ramps shown in top right panel. This is the closest we can get to the real astrophysical flux in this instance. For the orange and the violet-colored ramps the $\Delta DN$ values are anomalous until the green and red-colored ramps saturate. After that point, the orange and the violet-colored ramps appear to approximately follow a “classical” non-linearity, i.e., they show a drop in response. The picture becomes less clear in the PSF wings, for example in the violet and olive-colored ramps, as the latter deviate strongly despite the former only reaching 2/3rds of the pixel dynamic range. However, a 2D dependence of the observed effect has to be taken into account to get the full picture.

\section{Observations and MIRI BFE impact on science}
\label{sec:observations}

\subsection{The impact of the MIRI BFE on the imager PSF}
\label{subsec:bfe_imager}

\begin{figure}[t]
\includegraphics[width=0.48\textwidth]{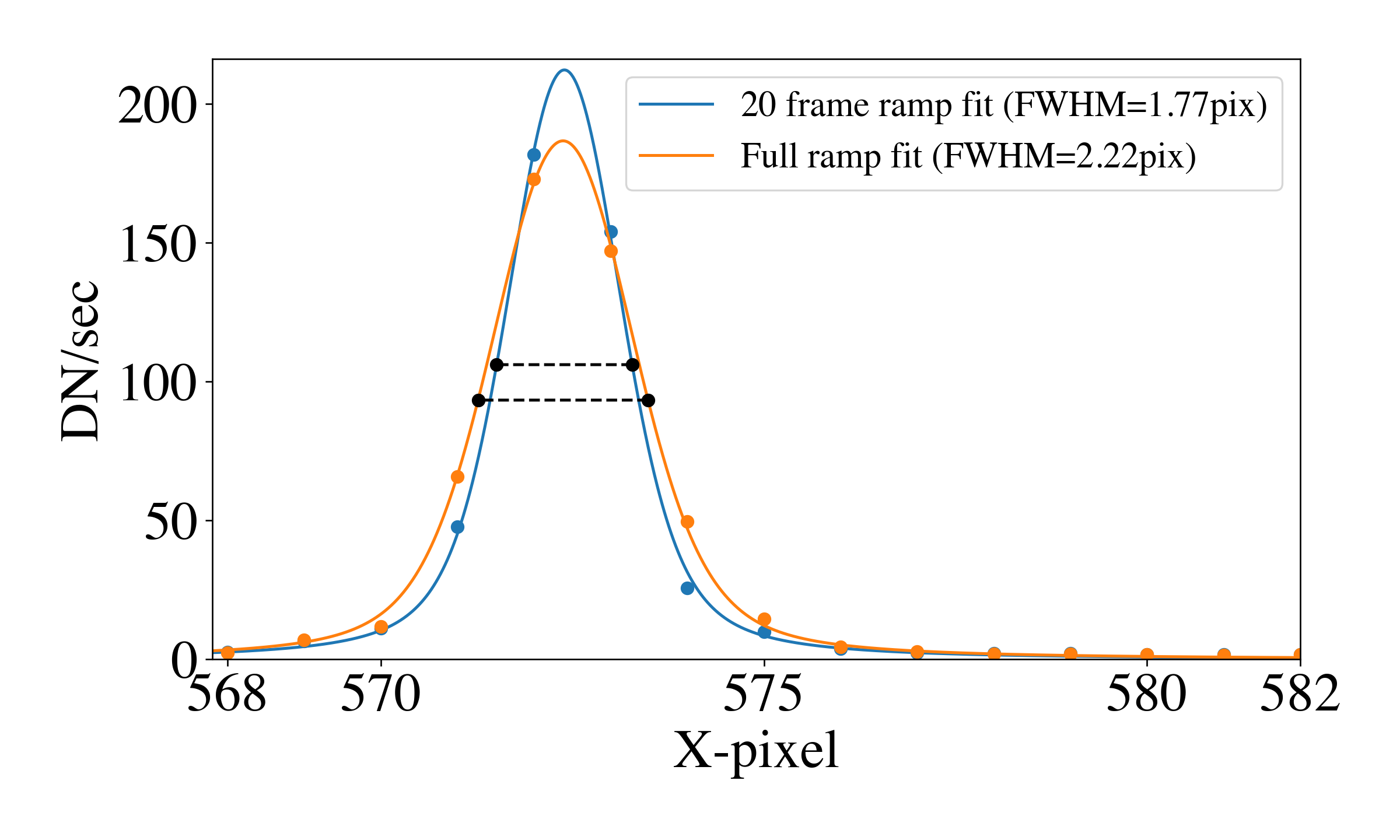}
\caption{MIRI imager PSF based on pixel ramp slope fits (20 frames versus full ramp). 1D Voigt profiles are fitted to the two PSFs and the FWHM is estimated based on where the fitted profiles have a signal value equal to half of the peak. The MIRI BFE results in a percentage change between the widths of the PSF of 25\% in the case where the full ramp is used for the slope fitting.}
\label{fig:bfe_miri_imager_impact}
\end{figure}

The data reduction workflow for \textit{JWST} data involves performing a linear regression on each of the pixel voltage integration ramps after these have been corrected for electronic effects; this includes the non-linearity correction \citep{morrison23}. If the impact of the MIRI BFE, as seen in Fig.~\ref{fig:bfe_signature_imager}, is not corrected or mitigated, the final estimated slope values from the linear regression (measured in DN per second) end up being systematically higher. The impact of this on the MIRI imager PSF is shown in Fig.~\ref{fig:bfe_miri_imager_impact}. Here we compare the case where, after flagging and omitting the saturated frames (frames whose signal is above 56.660~DN), and applying the \textit{JWST} MIRI non-linearity correction on the remaining frames, the slope values are manually computed from performing a linear regression on all the remaining frames in each ramp, versus only the first 20 frames in each ramp. We then fit a 1D Voigt profile\footnote{The fitting was performed using the \texttt{astropy.modeling} python package \citep{astropy:2013,astropy:2018}.} to the sampled imager PSF and estimate the profile Full-Width at Half-Maximum (FWHM) empirically based on where the fitted profiles have a signal value equal to half of the peak. The FWHM values of the two Voigt profile fits differ by 25\%. Conclusively, when using all the non-saturated frames in the ramps, as opposed to only a subset of frames at the start of the ramps, the resulting PSF appears fatter.

To mitigate the impact of the MIRI BFE, the \textit{JWST} pipeline flags the pixels that reach the saturation limit and then flags the frames of all surrounding pixels past the frame where the bright pixels saturate \citep{morrison23}. These flagged frames are then not used in the slope determination of the surrounding pixels. In the case of Fig.~\ref{fig:bfe_signature_imager}, the ramp of pixel X=572 saturates first at group number 56, hence for the ramps of pixels X=571 and X=573 only the part of the ramp up to frame number 56 is used to fit the slope. Performing the same exercise as for Fig.~\ref{fig:bfe_miri_imager_impact}, even with this mitigation strategy, the MIRI BFE still broadens the PSF by a factor of $\sim$10~\%. Furthermore, this mitigation strategy results in omitting usable frames from the slope fitting. Doing so necessarily results in a reduction of the final signal-to-noise ratio (S/N) due to having fewer samples in the ramps to fit.

There are two straight-forward ways that the MIRI BFE impacts the scientific interpretation of the MIRI imager data. Firstly, when it comes to astronomical imaging of point sources, astronomers perform aperture photometry by estimating the sum of the flux inside a circle with a radius that contains the largest fraction of the flux and minimizes the contribution of the background to said flux. Limiting this radius to the PSF core, aperture correction factors are applied to account for the fraction of the PSF present outside the aperture that is not accounted for. Given that the MIRI BFE broadens the PSF, this results in a systematic effect on the absolute flux calibration depending on the brightness of the source. Libralato et al., in prep., estimate a linear dependence of stellar brightness and estimated stellar magnitude caused by BFE using F560W imager data from commissioning PID \href{https://www.stsci.edu/jwst/science-execution/program-information.html?id=1464}{1464}. In a single exposure, with the imager field of view containing stars of different brightness levels, these stars were all observed with the same number of frames (the ramps have the same length). After flagging saturated frames and applying the non-linearity correction on the ramps, three sets of calibrations were performed. The first set used only the first 10 frames in the ramps for the ramp fitting, then  the photometric calibration was applied, and the brightness of the stars (in magnitudes) was measured using standard aperture photometry. The second and third set of calibrations used the first 50 frames and the first 125 frames for the ramp fitting, respectively, and then estimated the stellar brightness in the same way as for the 10 frame case. Comparing the 10 frame and the 50 frame case, there is a linearly increasing $\Delta$ from the fainter to the brighter stars going up to 0.03~mag. Comparing the 10 frame and the 125 frame case, the linearly increasing $\Delta$ goes up to 0.04~mag. 

Secondly, a different method to extract integrated fluxes from a point source is by performing PSF-weighted photometry. For this, an accurate model of the PSF is required so that the weights used for each pixel are representative of the detector illumination. Such a model PSF can be derived empirically and bright sources are preferred for this (Libralato et al., in prep.). Applying a model PSF based on bright source measurements, that are impacted by the MIRI BFE, on faint sources, which have the most to gain from an optimal flux extraction method such as PSF-weighted photometry, will result in the technique not meeting the theoretical noise-reduction predictions. Testing of PSF models is ongoing work, so the quantification of the impact of the BFE on PSF-weighted photometry and the resulting noise properties is part of future work.

\subsection{The impact of the MIRI BFE on the Low Resolution Spectrometer PSF}

\begin{figure*}[tp]
\centering
\includegraphics[width=0.9\textwidth]{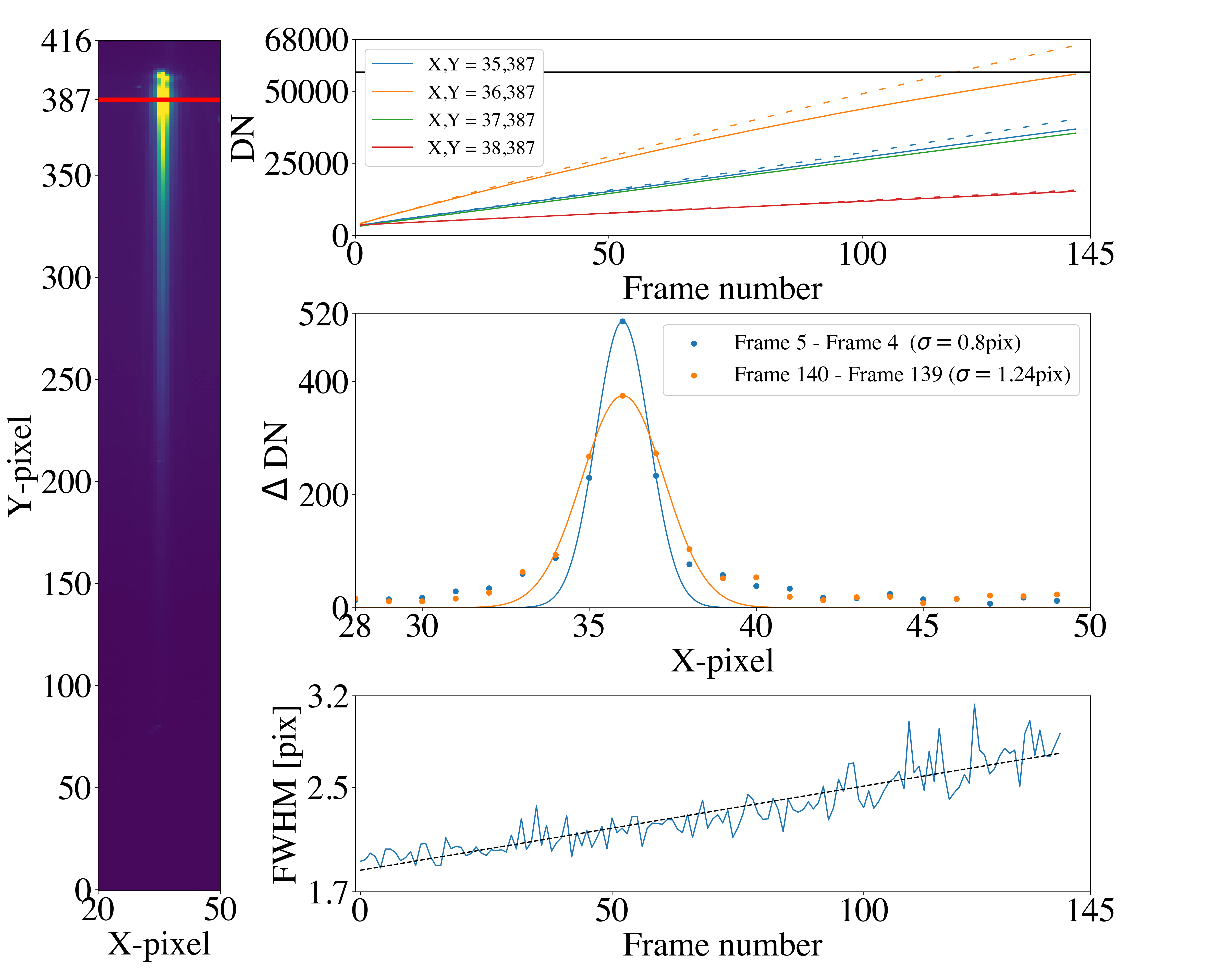}
\caption{MIRI LRS slitless prism data of an unresolved point source observed during the \textit{JWST} Cycle 1 phase. Left: One detector row is selected to examine how the MIRI BFE impacts the LRS PSF (red horizontal line). Right top: The solid lines are the raw pixel voltage integration ramps of the pixels sampling the PSF. The loosely dashed lines are the non-linearity corrected ramps. The horizontal black line at 56.600~DN shows the saturation value for the MIRI LRS pixels. Right center: Sampled PSF estimated from two sets of frame differences. 1D Voigt fits to the sampled PSFs are overplotted. Right bottom: LRS PSF FWHM estimated by taking subsequent frame differences along the length of the linearized ramps and fitting 1D Voigt profiles.}
\label{fig:bfe_miri_lrs_impact}
\end{figure*}

The MIRI Low Resolution Spectrometer (LRS) data are imaged on a subarray of the MIRI imager detector. During the \textit{JWST} Cycle 1 calibration phase a series of flux calibration activities was performed by observing flux calibration standards. In the left panel of Fig.~\ref{fig:bfe_miri_lrs_impact} we show the LRS spectrum of a time-series observation taken as part of PID \href{https://www.stsci.edu/jwst/science-execution/program-information.html?id=1536}{1536} of \href{http://simbad.cds.unistra.fr/simbad/sim-basic?Ident=BD%2B60+1753&submit=SIMBAD+search}{BD+60 1753}, which was observed with the LRS slitless prism. 

To study the impact of the MIRI BFE on the LRS PSF we use the recorded signal in a row of detector pixels that cross a bright part of the spectrum. In the top right panel of Fig.~\ref{fig:bfe_miri_lrs_impact} we show with solid lines four raw voltage integration ramps belonging to four of the pixels that sample the LRS PSF. We notice that none of the ramps reach saturation; this means that the BFE mitigation strategy currently implemented in the \textit{JWST} pipeline is not applicable here. The \textit{JWST} pipeline non-linearity corrected ramps are plotted as loosely dashed lines. The assumption is that, since these ramps have been non-linearity corrected, subsequent frame differences should yield the same mean slope if the ramps are indeed linear. In the middle right panel of Fig.~\ref{fig:bfe_miri_lrs_impact} we plot the LRS sampled PSF by taking two sets of frame differences, one set at the start of the non-linearity corrected ramps (loosely dashed lines), and one set at the end of the same ramps. 1D Voigt profiles are fitted to the two sampled PSFs. The FWHMs of the fitted Voigt profiles are significantly larger at the end of the ramp compared to the beginning of the ramp. This is quantified further in the lower right panel where the fitted PSF FWHM is seen to grow linearly along the ramp. This result implies that the ramps are not linearized (the loosely dashed lines are not linear); they show an increase in slope in the PSF wings and a decrease in slope in the PSF core. This is in line with the result shown previously for the MIRI imager in Fig.~\ref{fig:bfe_miri_imager_impact}.

In reality, each pixel on the left panel of Fig.~\ref{fig:bfe_miri_lrs_impact} recorded 212 integration ramps as part of the time-series observation of \href{http://simbad.cds.unistra.fr/simbad/sim-basic?Ident=BD%2B60+1753&submit=SIMBAD+search}{BD+60 1753} to characterize the photometric stability of the instrument. One of the major science cases for the LRS is observing exoplanet transits \citep{bouwman23}. These transits can sometimes result in up to a few percent change in the brightness of the host star depending on the relative star-to-planet size. Conceptually, this implies that for the integration ramps in transit, the ramps will reach a lower DN value compared to the integration ramps that are out of transit. As per the bottom right panel of Fig.~\ref{fig:bfe_miri_lrs_impact}, this will then result in a different PSF FWHM. For the LRS data of HD189733b (Program ID \href{https://www.stsci.edu/jwst/science-execution/program-information.html?id=2001}{2001}, PI: Michiel Min, Min et al., in prep), the impact on the transit spectrum (Rp/Rs)$^2$ is a transit depth that is 10~\% systematically too shallow at the shorter LRS wavelengths (5~$\mu m$) where the measured stellar spectrum is brightest.

\subsection{The impact of the MIRI BFE on Medium Resolution Spectrometer data of spectrally unresolved emission lines}

\begin{figure*}[htp!]
\centering
\includegraphics[width=0.8\textwidth]{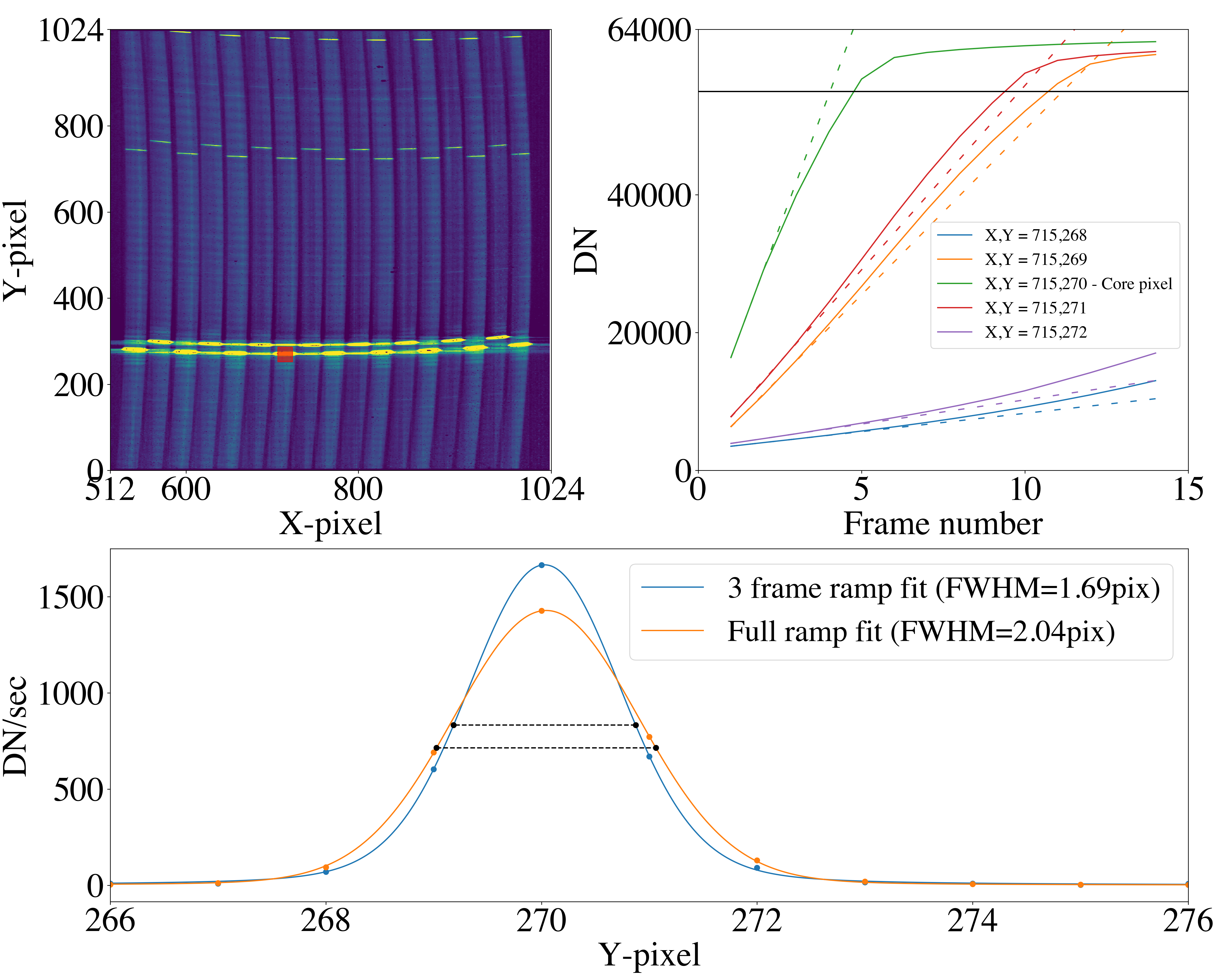}
\caption{MIRI MRS commissioning data of the spatially extended planetary nebula NGC 6543 (the Cat's Eye Nebula) in the 9.94--11.87~$\mu m$ range. Top left: Detector-level reduced MIRI MRS data. A small region of interest (red box) is selected on a bright emission line. Top right: Raw pixel voltage integration ramps of the pixels selected in the left image (solid colored lines). The dashed colored lines are produced from linear fits to the first 2 frames in each ramp. The horizontal black line at 55.000~DN shows the saturation value for the MIRI MRS pixels. Bottom: 1D Voigt fitted profile to the estimated emission line flux.}
\label{fig:bfe_signature_mrs}
\end{figure*}

In the case of MIRI observations of spectrally unresolved emission lines taken with the Medium Resolution Spectrometer (MRS), the line spread function (LSF) will be broadened in a similar fashion to the PSF shown in Fig.~\ref{fig:bfe_miri_imager_impact}. To illustrate this science case, in Fig.~\ref{fig:bfe_signature_mrs} we show the impact of the MIRI BFE on an unresolved emission line of the Cat's Eye Nebula \href{http://simbad.cds.unistra.fr/simbad/sim-basic?Ident=ngc+6543&submit=SIMBAD+search}{NGC~6543}, which was observed during the JWST commissioning phase (PID \href{https://www.stsci.edu/jwst/science-execution/program-information.html?id=1031}{1031}). The top left panel shows half of the MRS SW detector (columns 512 to 1024) covering the 9.94--11.87~$\mu m$ range. A number of curved strips in the vertical axis can be observed; these correspond to slits positioned on the sky of which the light is spectrally dispersed in the detector vertical direction along the strip curvature \citep{Wells_2015}. Since NGC~6543 is spatially resolved, all strips on the detector show flux from the spatially extended emission. We choose five pixels in the vertical direction covering the brightest emission line (location of red dot in the plotted panel). In the top right panel of Fig.~\ref{fig:bfe_signature_mrs} we show the shapes of the raw ramps as colored solid curves. The loosely dashed lines are lines that go through the first 2 frames in the ramps, and the horizontal black line shows the MRS saturation level at 55.000~DN. The raw ramps (solid) all visibly curve up above the loosely dashed lines due to the MIRI BFE, except for the one sampling the peak of the MRS LSF. The bottom panel of Fig.~\ref{fig:bfe_signature_mrs} shows the impact of the BFE on the determined shape of the emission line.

One recurring remark between the imager, LRS, and MRS data is that the signal in the core of the PSF/LSF, where the behavior of the non-linearity qualitatively follows the classical trend, also decreases at higher DN levels. This suggests that the non-linearity correction, derived from an extended illumination, does not perfectly linearize the ramps that sample the core of the PSF either. We discuss this further in Sect.~\ref{sec:discussion}.

\subsection{MRS spectral fringing and deriving an MRS non-linearity solution in the presence of BFE}

The MIRI MRS suffers from significant spectral fringing caused by Fabry-Pérot interference inside the MRS detectors \citep{argyriou2020SPIE,argyriou2020}. The fringing modulates the spectral baseline by up to 30\% of the continuum with a period between 12 and 30 pixels from 5~$\mu m$ to 28~$\mu m$. For point sources the fringe properties depend on (i) the profile of the MIRI pupil illumination, (ii) the part of the wavefront phase map that is sampled by the detector pixels, (iii) the wavelength of the in-falling light, (iv) the local geometry of the detector and (v) the refractive properties of the detector-constituting layers. For spatially extended sources the fringes depend only on (iii, iv, v). Nevertheless, this contrast of up to 30\% on the continuum results in a BFE-induced uncertainty on the derived non-linearity solutions. For MIRI the wavelength-dependent non-linearity solutions are computed empirically from measurements of the on-board internal calibration source flux \citep{morrison23}. Contrary to the MIRI imager where the fringes are blurred and the detector can be uniformly illuminated, for the MRS the fringes in the internal calibration source spectrum result in a BFE systematic on the non-linearity solutions driven by the fringing. After linearizing the ramps there is a residual 1$\sigma$ scatter of 3~\% (deviation from linearity), in contrast to the residual of 0.2~\% for the MIRI imager.

For the extended internal calibration source, the uncertainty on the global MRS non-linearity solution per spectral band is minimized by using the ramps of the pixels that sample only fringe peaks across the detector \citep{morrison23}. However, there is still a spread of $\sim$2~\% in the non-linearity solutions derived from the different pixels, and that is likely because lines of constant wavelength are curved on the MRS detector as noticed in the top left panel of Fig.~\ref{fig:bfe_signature_mrs}. This curvature changes across the surface of the detector. As such, pixels that sample fringe peaks interact differently with their surrounding neighbors depending on the exact distribution of flux in the neighbors.

\subsection{Implications of BFE on high contrast (coronographic) imaging}

A well characterized PSF is critical for high contrast imaging. As an example, \citet{Gaspar2023} found that exquisitely good PSF subtraction could be achieved at 25.5 $\mu$m using a reference star (\href{http://simbad.cds.unistra.fr/simbad/sim-basic?Ident=19+PsA&submit=SIMBAD+search}{19 PsA}) of very similar brightness to the science target (\href{http://simbad.cds.unistra.fr/simbad/sim-basic?Ident=Fomalhaut&submit=SIMBAD+search}{Fomalhaut}). Attempts to build and use a synthetic PSF using WebbPSF were far less successful. Until now the cause of the unsatisfactory performance of the synthetic images was attributed predominantly to the effect of detector scattering, which causes a cruciform in the imager and LRS PSF, and significant PSF broadening and excess power in the wings for the MRS PSF up to 12~$\mu m$ \citep{Gaspar2021,kendrew23,argyriou23}. The BFE is introducing an additional systematic on the PSF and LSF, one that is dependent on the covered dynamic range of the pixels. \citet{patapis2022} used MIRISim \citep{klaassen2021}, an instrument simulator with imaging based on WebbPSF, to evaluate the ability to obtain high quality MRS spectra of exoplanets. An understanding of the BFE is desirable to enhance the accuracy of such spectra and to allow obtaining them relatively close to the host stars. The ultimate performance of MIRI coronagraphy depends on utilization of a PSF reference star \citep{coronograph_perf}. Again, matching the star to the science target brightness is likely to improve the quality of this procedure. These issues can be mitigated by appropriate planning of observations, particularly given that the telescope shows high stability and wavefront drifts are small, even after large slews, providing a large selection of possible reference stars. Given that coronagraphic and high contrast imaging will use a limited number of spectral bands, additional flexibility may be possible by building empirical models of the image as a function of brightness. 

\begin{table*}[htp!]
\caption{MIRI Si:As IBC detector electrostatic model parameters inputted to \texttt{Poisson\_CCD}.}
\centering
\begin{tabular}{lll}
\hline\hline
Parameter & Value [unit] & Description \\
\hline
\texttt{V\_{bb}}   & -4.0~V  & Back bias voltage at buried contact \citep{miri_pasp_8}  \\
\texttt{V\_{contact}}   & -1.8~V  & Contact voltage at frontside contact plus 0.2~V from clock feedthrough \\
\texttt{ContactCapacitance} & 33.6~fF  & Nominal capacitance at the integrating node \citep{miri_pasp_7} \\
\texttt{RecombinationLifetime}  & $2\,.\,10^{-7}$~sec & Electron recombination time \citep{miri_pasp_7}\\
\hline
\texttt{TopDopingThickness}  &  1.0~$\mu m$  & Thickness of buried contact \\
\texttt{TopSurfaceDoping} & $7.5\,.\,10^{18}$~cm$^{-3}$  & Doping buried contact \citep{argyriou2020SPIE} \\
\texttt{BackgroundDoping} & $1.5\,.\,10^{12}$~cm$^{-3}$ & Density of minority impurity concentration in infrared-active layer \\
\texttt{BottomOxide}  & 1.0~$\mu m$   &  Bottom oxide thickness above pixel metallization \\
\texttt{ContactDose\_0} & $7.5\,.\,10^{18}$~cm$^{-3}$  &  Doping of frontside contact \\
\texttt{ContactWidth} & 20.0~$\mu m$  &  Width of frontside contact \\
\texttt{ContactHeight} & 20.0~$\mu m$  &  Height (not thickness) of frontside contact \\
\hline
\texttt{SensorThickness}   & 35~$\mu m$  & Infrared-active layer thickness  \\
\texttt{PixelSizeX}   & 25~$\mu m$  & Pixel size in x \citep{miri_pasp_7}  \\
\texttt{PixelSizeY}   & 25~$\mu m$  & Pixel size in y  \\
\hline
\texttt{CCDTemperature} & 6.4~K & Operational temperature of MIRI detectors \\
\texttt{Lambda} & 5.6~$\mu m$ & Wavelength of incoming light \\
\hline
\label{tab:miri_poisson_ccd}
\end{tabular}
\vspace{-6mm}
\end{table*}

\begin{figure*}[htbp!]
\centering
\includegraphics[width=0.9\textwidth]{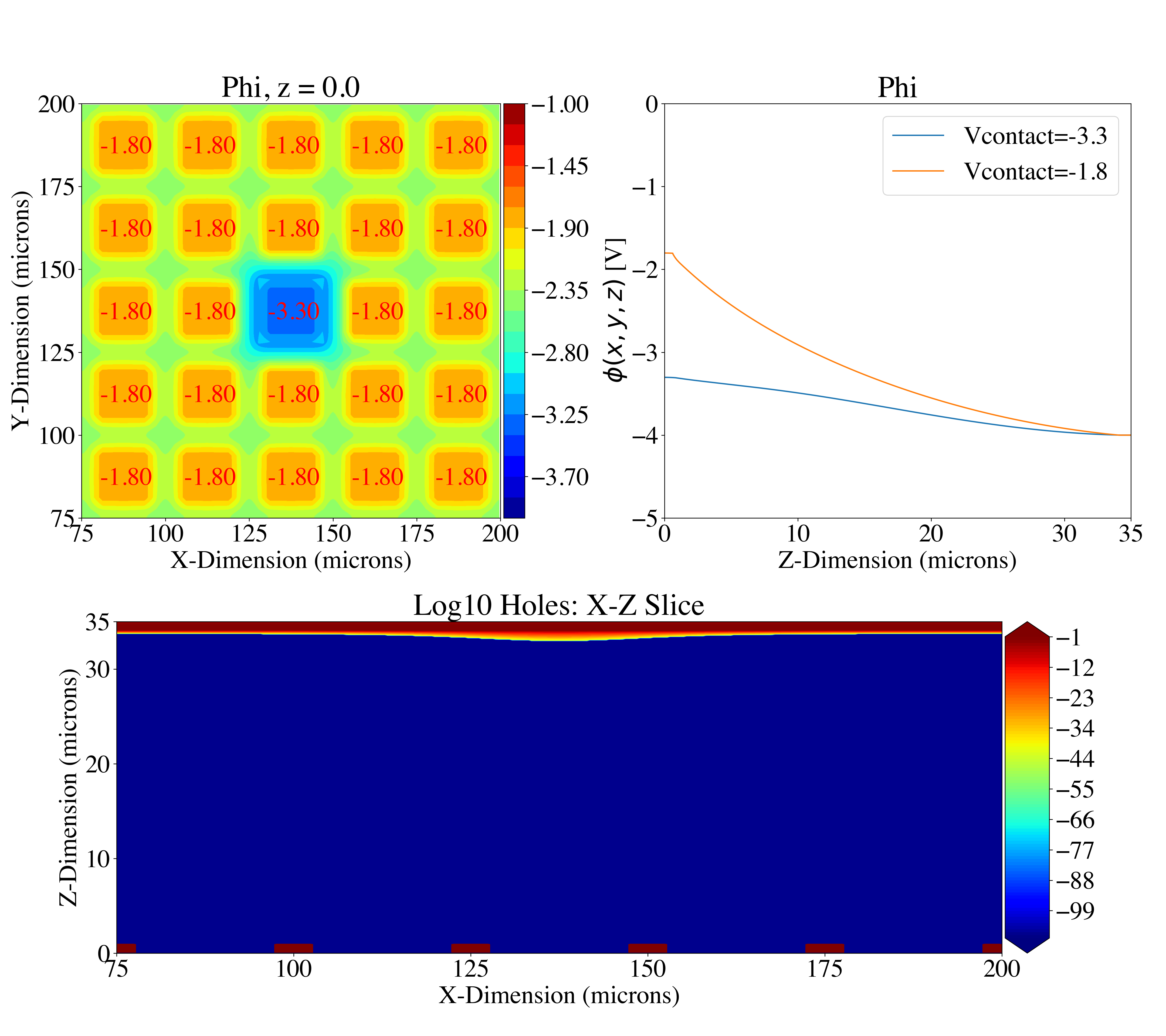}
\caption{\texttt{Poisson\_CCD} output plot of basic MIRI detector model simulation. Top left: Poisson's equation is solved for a 5-by-5 grid of pixels where a localized voltage debias of $-1.5~V$ is applied at the middle pixel. Top right: The electrostatic potential is shown as a function of detector depth. Z=0~$\mu m$ is located at the pixel, Z=35~$\mu m$ is located at the top of the buried contact. Bottom: The MIRI detector infrared-active layer. The blue shading shows areas where the active layer is fully depleted. Yellow and red shading show areas where the active layer is undepleted. The buried contact is located at Z=34--35~$\mu m$. The dark red boxes at Z=0--1~$\mu m$ illustrate the regions not covered by the detector frontside contact.}
\label{fig:poisson_ccd_pixelsim}
\end{figure*}

\section{Modeling}
\label{sec:modeling}

\subsection{Setting up the MIRI Si:As IBC detector array electrostatic model using \texttt{Poisson\_CCD}}

The public code \texttt{Poisson\_CCD}\footnote{\url{https://github.com/craiglagegit/Poisson_CCD/tree/master}} solves Poisson’s electrostatic differential equation given by Eq.~\ref{eq:poisson_equation} numerically, in 3 spatial dimensions, and simulates charge transport within CCDs \citep{Lage2021}. The potentials and free carrier densities within a CCD are self-consistently solved for, giving realistic results for the depletion profile and electron paths followed in silicon. 

\begin{equation}\label{eq:poisson_equation}
    \nabla^2 \phi(x,y,z) = \frac{\rho(x,y,z)}{\epsilon_{Si}}
,\end{equation}

\noindent where $\phi$ is the electrostatic potential, $\rho$ is the charge density, and $\epsilon_{Si}$ is the dielectric constant of silicon. 

Contrary to CCDs, where the entire silicon substrate needs to be accounted for, in the case of the Si:As IBC detectors only the narrower infrared-active layer needs to be modeled. For this, an input file is defined with all necessary parameters to solve Eq.~\ref{eq:poisson_equation} numerically. The MIRI-specific parameters are tabulated in Table~\ref{tab:miri_poisson_ccd}, where some of these parameters were previously mentioned in Sect.~\ref{subsec:pixel_readout} and \ref{subsec:miri_nonlinearity}. The doping level of the contacts \texttt{ContactDose\_0} of N=$7.5\,.\,10^{18}$~cm$^{-3}$ was estimated in \citet{argyriou2020SPIE} by modelling the high and low frequency fringing present in MRS data\footnote{In case a smooth Gaussian profile is preferable for the doping distribution across the contacts the \texttt{Poisson\_CCD} parameters \texttt{ContactDose\_0}, \texttt{ContactPeak\_0}, and \texttt{ContactSigma\_0} can be used to produce the required integrated doping level}. The parameter \texttt{Lambda} is used in conjunction with Eq.~\ref{eq:absorption_coefficient}, implemented into the \texttt{Poisson\_CCD} code, to estimate at which depth (Z-axis) in the infrared-active layer each photoelectron is produced.

One of the simplest simulation setups in \texttt{Poisson\_CCD} is shown in Fig.~\ref{fig:poisson_ccd_pixelsim} for MIRI. A 5-by-5 pixel grid is used, as seen in the top left panel staring down at the detector pixels. The frontside contact voltage is set at the value of \texttt{V\_contact} and the buried contact voltage is set by \texttt{V\_bb} defined in Table~\ref{tab:miri_poisson_ccd}. A localized debiasing \texttt{DeltaV\_0\_0}=-1.5~V is applied at the center pixel. The variation of the electrostatic potential $\phi(x,y,z)$ across the depth of the infrared-active layer is shown in the top right panel. Finally the concentration of holes in the infrared-active layer is shown in the bottom panel. The blue shading illustrates where the active layer is depleted, while the yellow and red shading denotes areas where the layer is undepleted. In this simple case, we see a marginal reduction in the depletion region width in the central pixel compared to the neighboring pixels where the infrared-active layer is fully depleted. The top 1~$\mu m$ (dark red shading) is the location of the buried contact. The 1~$\mu m$-thick dark red shading at Z=0~$\mu m$ is the location not covered by the frontside contact, which itself rests on the pixel metallization as illustrated in Fig.~\ref{fig:detector_layout}.

\subsection{Electrostatic simulation for MIRI imaging case}
\label{subsec:electrostatic_simulation}
To perform a realistic MIRI simulation of a point source illumination on the detector, the \texttt{Poisson\_CCD} code is run with a random set of tracers (photons) pooled from a 2D Gaussian pattern. The Gaussian pattern 2D widths in x and y-directions are defined in Table~\ref{tab:miri_poisson_ccd} based on a 2D Gaussian to the 20 frame ramp fit PSF in Fig.~\ref{fig:bfe_signature_imager}. We find $\sigma_{x}$= 0.84 pix, $\sigma_{y}$=0.82~pix, equivalent to \texttt{Sigmax}~=~20.9~$\mu m$ and \texttt{Sigmay}~=~20.5~$\mu m$ used in \texttt{Poisson\_CCD}. We also input the measured sub-pixel centroid from the same 2D Gaussian fit (parameters \texttt{Xoffset}~=~10.75~~$\mu m$, \texttt{Yoffset}~=~9.75~$\mu m$) as this has an effect on the relative signal level between neighboring pixels, as exemplified in the bottom left panel of Fig.~\ref{fig:bfe_signature_imager}. 

\texttt{Poisson\_CCD} first solves Eq.~\ref{eq:poisson_equation}, then a distinct number of tracers are injected into the model until the hardcoded number of detected and recombined photoelectrons, defined by the parameter \texttt{NumElec}, is reached. Every time a tracer results in a photoelectron, produced at a given pixel based on the absorption length of Eq.~\ref{eq:absorption_coefficient}, the path of the photoelectron, dictated by the electric field inside the layer, is numerically computed and the pixel at which the photoelectron ends up in, is recorded. The photoelectric charge accumulation introduces a debiasing at each pixel as defined by the \texttt{ContactCapacitance} parameter. The updated debiasing level at each pixel is used to solve Eq.~\ref{eq:poisson_equation} again in the next iterative step. Performing such a simulation self-consistently results in each pixel shown in the top left panel of Fig.~\ref{fig:poisson_ccd_pixelsim} experiencing a different level of debiasing. Pixels that see more tracer photons will have more photoelectrons generated above them, increasing the resulting debiasing, and reducing the width of the depletion region as a result. The impact of the debiasing on the path of the photoelectrons is monitored by tracking the electron paths in the active layer. The number of photoelectrons lost due to recombination is computed as the number of tracers injected minus the number of detected photoelectrons at the end of each simulation step.

\begin{figure*}[htp]
\centering
\includegraphics[width=0.88\textwidth]{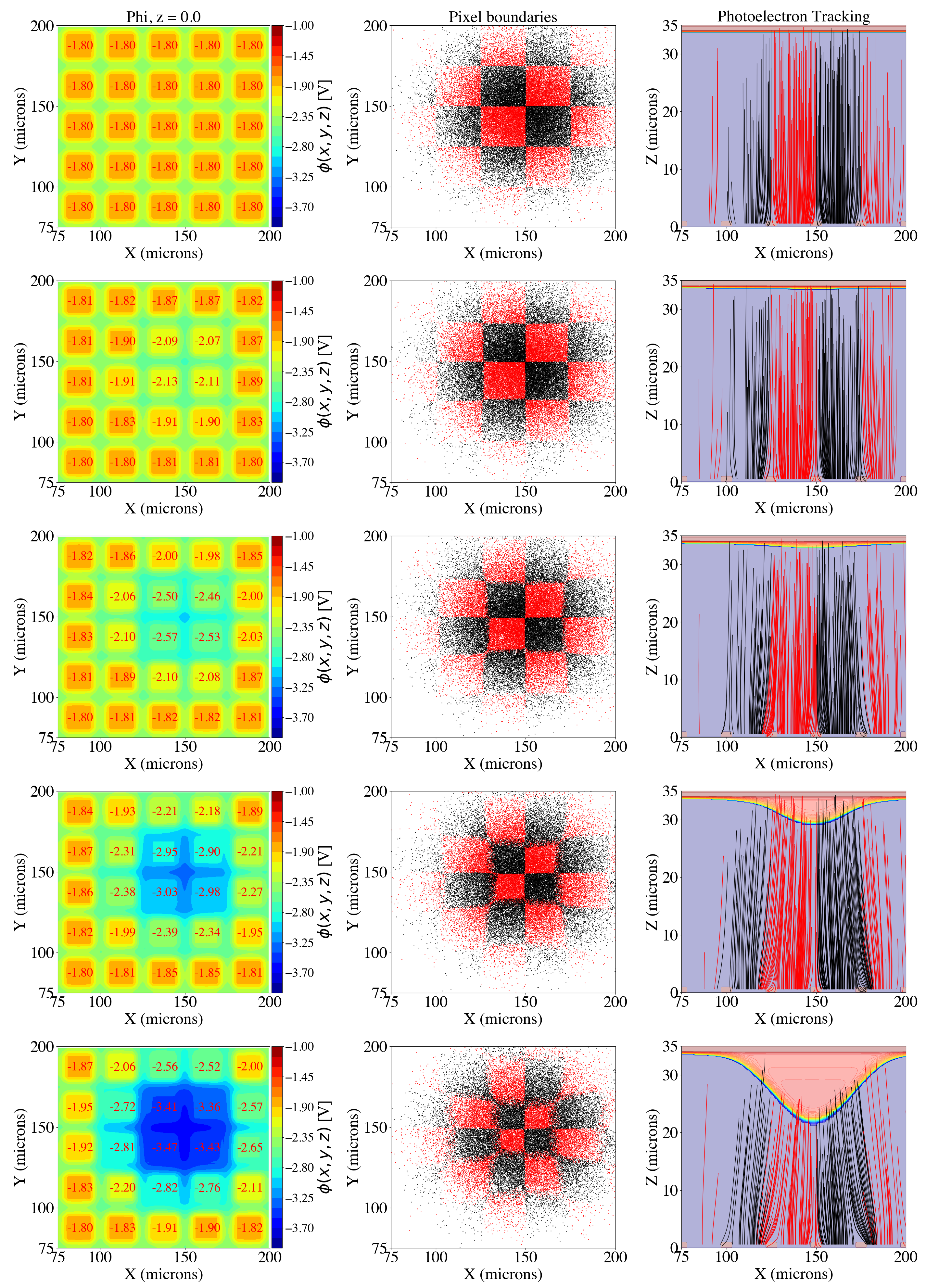}
\caption{Brighter-Fatter Effect in MIRI detector model simulation. Left column: Distribution of the electrostatic potential $\phi$ across the 5-by-5 simulation pixel grid. Middle column: Location on the pixel geometric grid where each produced photoelectron landed. Neighboring pixels are given alternating colors. This probes the effective pixel collecting area change due to the debiasing. Right column: Photoelectron generation in infrared-active layer and photoelectron path tracking. The light pink region is the undepleted region, which grows as the pixel accumulates more charge. Going from the top to the bottom row, the model is injected with 1, 10, 25, 45, and 75 times N=45,000 tracers. }
\label{fig:poisson_ccd_bf_tests}
\end{figure*}

In Fig.~\ref{fig:poisson_ccd_bf_tests}, we show five simulation run steps, specifically, run number 0, 10, 25, 45, and 75. \texttt{NumElec}=45,000 photoelectrons are produced in the model at each step. The photoelectrons are distributed over a 2D Gaussian profile. In total we ran the simulation with 125 steps, which took 50 hours (2 days) real time on a single core of a Macbook Pro with a 2.3 GHz Intel Core i9 processor and 16 GB of RAM. The value of \texttt{NumElec} was defined such that we would get a similar $\Delta DN$ level as the green ramp in the bottom right panel of Fig.~\ref{fig:bfe_signature_imager} (the value of \texttt{NumElec} was determined by manual inspection; running an optimization using the simulations was too costly in terms of time). In the left column of plots in Fig.~\ref{fig:poisson_ccd_bf_tests}, the level of debiasing at each pixel is shown. In the middle column we show where each produced photoelectron lands on the geometric grid. In the right column the photoelectron paths are tracked from their creation site down to the pixels (the code does this in both x- and y-directions but we only plot the paths in the detector x-direction). The paths are overplotted on the concentration of holes in the infrared-active layer, which probes the areas of depletion and undepletion in the layer. The rows of plots go from lower debiasing in the top row to higher debiasing in the bottom row.

Due to the larger accumulation of photoelectrons in the central pixel, it records a larger debiasing level compared to its neighboring pixels. The \texttt{Poisson\_CCD} code is able to self-consistently update the debiasing across the 5-by-5 pixel grid. In the middle column of Fig.~\ref{fig:poisson_ccd_bf_tests} a decrease in the area of the (red) central pixel can be noticed from the top to the bottom row. This reduction in effective photocollecting area is explained by the plots in the right column. The panel in the bottom row shows clearly that photoelectrons produced above the central pixel are guided by the electric field to the neighboring pixels instead. In addition, in the same panel it can be noticed that a significant portion of the infrared-active layer is undepleted, and photoelectrons produced in that area recombine after a time span of \texttt{RecombinationLifetime} and are effectively "lost", i.e. they are not detected. In Table~\ref{tab:some_numbers} we provide the number of photoelectrons detected in grid pixel [2,2], as well as the accumulated DN value (a conversion factor of 5.5~e$^-$/DN is applied). We mention again that the \texttt{Poisson\_CCD} simulations are such that the detected and recombined photoelectrons at each run step, for all grid pixels, distributed over the sampled 2D Gaussian profile, is equal to \texttt{NumElec}=45,000.

\begin{table}[h]
\caption{Number of detector photoelectrons in pixel [2,2] (0-index) of the simulation output shown in Fig.~\ref{fig:poisson_ccd_bf_tests}.}
\centering
\begin{tabular}{lll}
\hline\hline
Run number & Photo-e$^-$ detected & Accumulated DN  \\
\hline
0   & 7,316  & 1,330   \\
10  & 76,637  & 13,934 \\
25 & 167,252  & 30,409 \\
45  & 262,880 & 47,796 \\
75  & 353,887 & 64,343 \\
\hline
\label{tab:some_numbers}
\end{tabular}
\vspace{-6mm}
\end{table}

\begin{figure*}[t]
\centering
\includegraphics[width=0.98\textwidth]{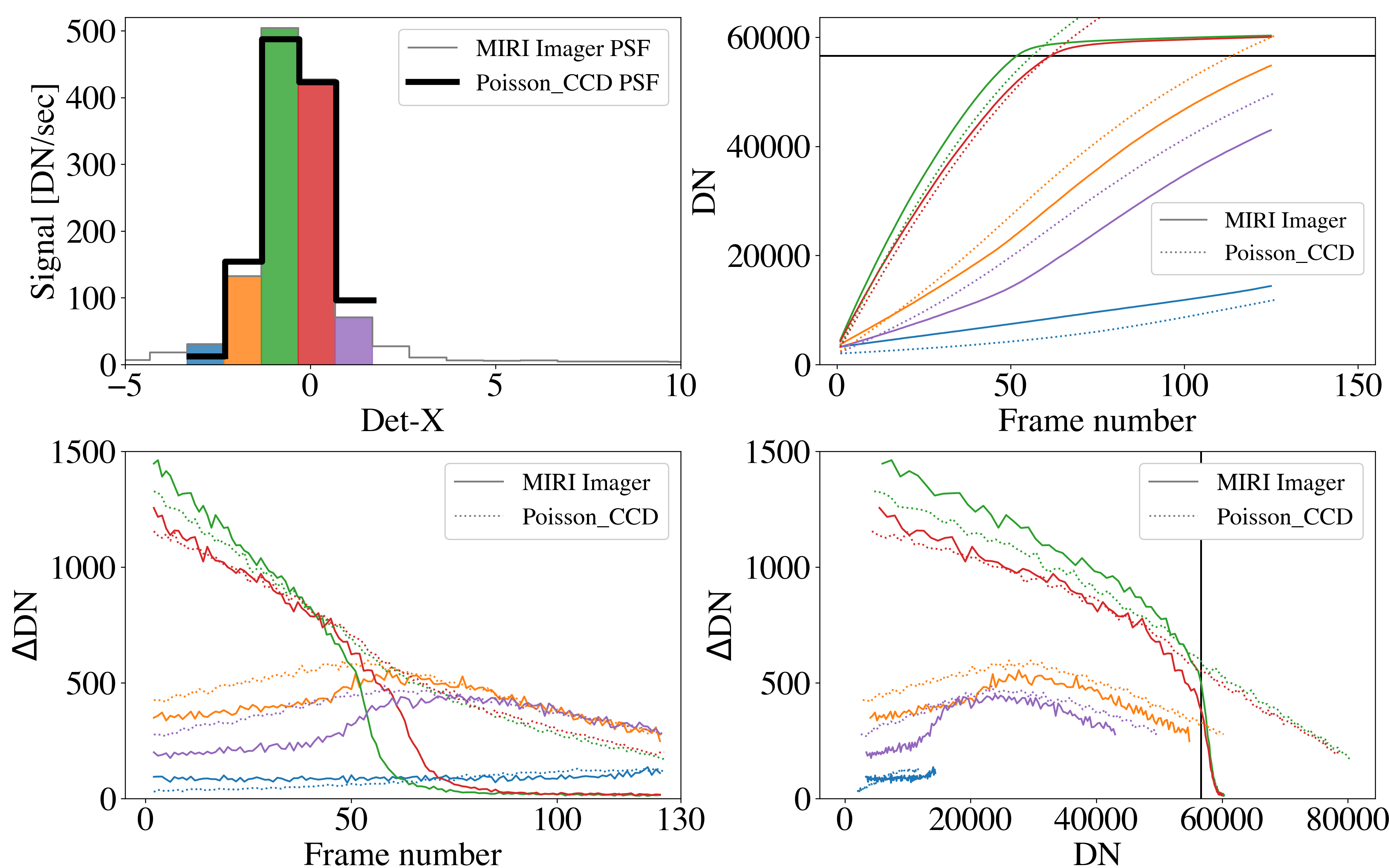}
\caption{Modeling the MIRI detector pixel integration ramps using Poisson\_CCD. Top left: Signal distribution in MIRI imager PSF in 1D versus the \texttt{Poisson\_CCD} simulated PSF (2D Gaussian). Both PSFs are centered on the x-axis using their respective centroids. Top right: MIRI imager raw ramps (solid colored curves) and ramps from the \texttt{Poisson\_CCD} simulation output (dotted lines). The frame number in the case of the \texttt{Poisson\_CCD} simulation corresponds to how many times N=45.000 tracers have been added to the simulation. To compute the DN level for the simulated pixels, the number of electrons that reached each pixel in each simulation step was divided by the pixel gain of 5.5 e-/DN. For visualization purposes we offset the \texttt{Poisson\_CCD} vertically by 2,000~DN. Bottom left: Subsequent frame differences as a function of time (frame number). Bottom right: Subsequent frame difference as a function of the pixel absolute DN level. The vertical black line shows the saturation limit.}
\label{fig:poisson_ccd_ramps}
\end{figure*}

Figure~\ref{fig:poisson_ccd_ramps} shows the comparison between the MIRI imager data shown in Fig.~\ref{fig:bfe_signature_imager} and the \texttt{Poisson\_CCD} simulation. In the top left panel we show the same MIRI imager PSF in 1D as what was shown in the bottom left panel of Fig.~\ref{fig:bfe_signature_imager}. We overplot the 1D signal distribution of the 2D Gaussian PSF of the \texttt{Poisson\_CCD} simulation output in run step 1. We can see that the relative signals in the MIRI and the \texttt{Poisson\_CCD} case are similar but do not match perfectly. That is due to the small mismatch in sampled PSF shapes (the shape of the MIRI PSF is not Gaussian) and the uncertainty on the fitted 2D centroid and widths. In the future it will likely be possible to input a custom PSF into the \texttt{Poisson\_CCD} code, however, this is not currently the case. In the top right and bottom left panels of Fig.~\ref{fig:poisson_ccd_ramps} we show the same MIRI imager raw ramps in absolute DN and $\Delta$DN space from the top right and bottom right panel of Fig.~\ref{fig:bfe_signature_imager} as solid lines, as well as the results from the \texttt{Poisson\_CCD} simulation as dotted lines. For the simulated ramps, the DN level was computed from the number of photoelectrons landing on each pixel, divided by the pixel gain of 5.5~e$^-$/DN \citep{miri_pasp_7}. Since it is not possible to realistically simulate saturation for the MIRI pixels, the ramps cross the saturation limit unhindered. In the bottom right panel of Fig.~\ref{fig:poisson_ccd_ramps} we plot the $\Delta$DN values against the absolute DN level. It is in this time-independent space that the MIRI non-linearity calibration solution is derived for all pixels, as discussed in detail in \citet{morrison23}.

Studying Fig.~\ref{fig:poisson_ccd_ramps}, we can quantify the differences between the MIRI data and the simulations in three ways, (1) the differences in the modeled PSFs in the top left panel, (2) the differences in the absolute DN and $\Delta$DN levels of the real and simulated ramps in the top right and bottom left panels, (3) the profile of the curvatures in the bottom right panel. Regarding (1), subtracting the PSF values and computing the residuals in a standard deviation sense is not useful when discussing the BFE. What matters is the absolute level of each pixel, and that in relation to the level of the neighboring pixels. In Table~\ref{tab:comparison} we provide statistics that play a role in reproducing the BFE that manifests in the MIRI ramps. We find the largest deviations in the PSF wings, which is indeed where the MIRI PSF becomes least Gaussian. Nevertheless this location is of interest for many science aspects of MIRI, some of which were presented in Sect.~\ref{sec:observations}. Consequently, for (2) the results are immediately linked to the discussion of (1). In absolute DN space, the injected number of tracers in \texttt{Poisson\_CCD} dictates the slope of the ramps, and this can be optimized once the correct PSF shape is used. Qualitatively the simulations in the top right and bottom left panels of Fig.~\ref{fig:poisson_ccd_ramps} reproduce the non-linearity observed in the PSF peak (green and red ramps), linked to the reduction in the depletion region width seen in the right column of Fig.~\ref{fig:poisson_ccd_bf_tests}. Off the PSF peak (blue, orange, violet ramps) the simulations match the observations less well in the time (frame number) domain. To focus on the flux and time-independent behavior of the ramps, the bottom right panel is most applicable. Here the $\Delta$DN is plotted as a function of DN on the x-axis. For a spatially uniform illumination of the detectors, all the pixels show a non-linear trend as per the green solid curve. Interestingly, we see that there is a small angle between the green and red ramps, both for MIRI and the \texttt{Poisson\_CCD} results. This suggests that even a small illumination contrast can result in a BFE systematic. For the green ramp we measure a maximum of 8~\% difference between observations and simulations. This number drops to 3.5~\% for the red ramp, with this difference being located in the low DN regime; the difference reduces to a sub-percent level above 25,000~DN. This positive result can be linked back to the statistics shown in Table~\ref{tab:comparison}, where the red pixel is where the best match between the simulated and observed PSF was achieved. For the orange and violet ramps, the match with the observations is better above 30,000~DN and 20,000~DN respectively, with an offset of 10~\%, however, the match is less good below those DN values, with a maximum offset of 60~\%. We find that the location of the maximum in the curves is a positive indication that the simulations reproduce the impact of the BFE on the observed ramp slope change as a function of DN. We strongly believe that using the MIRI PSF will improve the match for all pixels, also for the fainter blue pixel, although as we will discuss next, using a self-calibration approach, constrained by our findings, is likely to be a more efficient approach to correcting the ramps, given the computational power required for these simulations.

\begin{table}[h]
\caption{Statistics on the MIRI vs simulated PSFs.}
\centering
\begin{tabular}{llll}
\hline\hline
Pixel color & $\left( \frac{pix[i]_{MIRI}}{pix[i]_{sim}} - 1 \right)$ & $\left( \frac{pix[i]}{pix[i-1]}\right)_{MIRI}$ &  $\left( \frac{pix[i]}{pix[i-1]}\right)_{sim}$ \\
\hline
blue   & 187\%  & - & -  \\
orange  & 16\%  & 5.0 & 12.5 \\
green & 3.9\%  & 3.3 & 3.7 \\
red  & 0.9\% & 0.85 & 0.87 \\
violet & 34.7\% & 0.22 & 0.17 \\
\hline
\label{tab:comparison}
\end{tabular}
\vspace{-6mm}
\end{table}

\section{Discussion}
\label{sec:discussion}

The results shown in Fig.~\ref{fig:poisson_ccd_bf_tests} and \ref{fig:poisson_ccd_ramps} converge to four important outcomes: (1) They provide strong evidence for the physical mechanism driving the BFE for point sources (and any source with a flux contrast on the detector). The MIRI BFE is caused by the detector debiasing altering the configuration of the 3D electrostatic potential $\phi(x,y,z)$ inside the Si:As IBC sensor infrared-active layer. The corresponding electric field guides photoelectrons away from the pixel above which they were generated into neighboring pixels. Linking this back to the ramp plots shown in the right panel of Fig.~\ref{fig:bfe_signature_imager} and the top right panel of Fig.~\ref{fig:bfe_signature_mrs}, at higher DN levels the debiasing in the central pixel results in an increasing fraction of photoelectrons being detected at the neighboring pixels instead, causing an increase in the measured ramp slopes of the neighboring pixels. Furthermore, this effect can extend to the pixels not directly neighboring the central pixel, as those that do neighbor the central pixel also start contributing to their own neighbors as seen in the bottom two panels of the right column in Fig.~\ref{fig:poisson_ccd_bf_tests} and the right panel of Fig.~\ref{fig:poisson_ccd_ramps}. (2) The increasing size of the active layer undepleted region increases the number of photoelectrons lost via recombination. This is the physical mechanism driving the MIRI non-linearity for a homogeneous (spatially uniform) detector illumination. Interestingly, even though at higher debiasing levels the photoelectrons that are detected start their path closer to the pixels, these are still guided to the neighboring pixels by the electric field. (3) The profile of the undepletion region contour within each pixel is not flat. The profile shows a significant gradient over the span of a pixel width. This implies that (a) the center pixel will have a non-linearity solution akin to that of a spatially uniform (extended) source, however, (b) the photoelectron collection profile likely has an intra-pixel dependence on where the centroid of the source is. This means that even the brightest pixel will have a different non-linearity calibration solution depending on the point source centroid. (4) Perhaps most important, based on the previous three points, it is clear that the MIRI detector non-linearity and BFE cannot be considered to be independent of one another in terms of calibrating the two effects. A JWST Cycle 2 calibration proposal (PIs: Danny Gasman and Ioannis Argyriou, PID \href{https://www.stsci.edu/jwst/science-execution/program-information.html?id=3779}{3779}) has been proposed and accepted to study this issue further by performing a 3-by-3 intra-pixel raster scan around the 8 nominal point source MRS dither positions.

In Sect.~\ref{sec:introduction} we mentioned that, contrary to CCDs and Hawaii-XRGs (HgCdTe) detectors, there is a fundamental question of whether the BFE can manifest in Si:As IBC devices at all. We find in this paper that, although the manifestations of the BFE and the impact on the data are similar for the three detector types, the mechanisms behind it have a number of distinct differences. For CCDs, the BFE arises when photoelectrons from a large signal accumulate at a gate and partially shield the detector volume from the electric field for that gate. This effect is direct because the accumulated charge remains within the detector volume. The result is that the field lines are distorted in a way that reduces the effective area of the pixel as viewed by a photo-excited electron created near the entrance surface of the detector. No signal is lost, but photoelectrons under the influence of this modified field are directed to pixels surrounding the one with the large accumulation of photoelectrons. In addition, it is important to consider that a typical CCD has a large backside bias (10~V in the context of the \textit{Euclid} space telescope \citep{clarke2015}, and 45~V in the context of the ground-based Large Synoptic Survey Telescope LSST \citep{chun2010}), keeping the substrate fully depleted throughout an exposure. So even though the accumulated charge increases the voltage on the storage node by 1-2V, the substrate remains fully depleted. The primary cause of the BFE in this case is the change of the electric field at the pixel, resulting in the repulsion of incoming carriers. For photodiode arrays, e.g. the H2RG devices used by the \textit{JWST} NIRISS, NIRCam, and NIRSpec instruments, a depletion region at the diode junction provides the large impedance needed for low noise performance and more importantly for our discussion, it enables the diffusion current and charge collection. The width of this region is reduced for pixels with large accumulated signals, which lowers the diffusion-induced current. The result is that neighboring pixels can attract photoelectrons through their depletion regions that are in the undepleted volume over this pixel and hence have no significant electric field to drive them across the junction. Again, the primary effect is to redirect photoelectrons to neighboring pixels without significant losses. For Si:As IBC devices, the high impedance is provided by the low temperature of operation and the blocking layer; the generation of photoelectrons occurs within the depletion region, and for high quantum efficiency this layer must be an order of magnitude larger than in photodiodes, e.g., $\sim$~30~$\mu m$. In this case, a large charge on a pixel reduces its bias voltage and causes its depletion region to retreat toward the output (front) contact. The outcome can be an undepleted column under this pixel of length $\sim$~10~$\mu m$ or more, surrounded by depleted columns for the surrounding pixels, and an electrostatic formalism similar to that developed for CCDs can be used to model the resulting apparent shrinkage of the high-signal column as viewed by photoelectrons entering the backside of the detector. Unlike CCDs and photodiodes, for Si:As IBC detector arrays a significant fraction of photoelectrons are lost when they recombine in the undepleted column. As a result, although CCDs and photodiodes can exhibit blooming with very large signals, where the excess photoelectrons migrate in bulk to surrounding pixels, Si:As IBC devices do not show this behavior (Dicken et al., 2023, in prep.). Another difference is that, because the absorption coefficient for Si:As IBC material has a strong wavelength dependence, the BFE is likely to be subtly wavelength dependent.

The \texttt{Poisson\_CCD} model presented in this paper could in theory be used to derive a simultaneous correction for the MIRI BFE and non-linearity, however, there are several limitations to consider. (1) In the case of MIRI MRS and LRS spectroscopy, such an endeavor is extremely more complex given that the spectral continuum is dispersed in curved strips on the detector covering thousands of pixels, as shown for example in the top left panel of Fig.~\ref{fig:bfe_signature_mrs} for the MRS. (2) The computational time for one simulation is significant, and given that MIRI observations are dithered, this means that several simulations would be required for one observation. Since there is uncertainty linked to the centroid of the source in the sub-pixel level (linked to pointing jitter, guide star information, and pointing accuracy with and without target acquisition) and this has a significant effect on the results (the pixel phase has to be correct); introducing a centroid optimization loop would be advisable, but would also increase the computational cost.

We discuss the possibility of using a self-calibration approach, applied directly from, and to, the raw ramps. By re-injecting the charges of the neighboring pixels back where they should have been measured, the effect of the BFE on the ramps can be minimized. The mathematical operation is equivalent to a deconvolution. Importantly, it can be applied to all operational modes of MIRI, i.e., to both imaging and spectroscopy. In Fig.~\ref{fig:poisson_ccd_bf_tests} we showed, based on our electrostatic simulations, that the furthest a photoelectron can travel is one pixel adjacent. This is useful knowledge when deriving a deconvolution kernel. It means that the kernel can be limited to a 3-by-3 matrix of values. The matrix values can be computed at each frame number based on the absolute DN level of each pixel, which defines the level of debiasing and width of the depletion region. This work falls outside the scope of this first paper and is presented in Gasman et al., Paper II, in prep.

The JWST Cycle 2 calibration proposal PID \href{https://www.stsci.edu/jwst/science-execution/program-information.html?id=3779}{3779} mentioned in Sect.~\ref{subsec:electrostatic_simulation} will provide essential information for furthering the work on the MIRI MRS BFE characterization and calibration. In this calibration program the absolute flux standard \href{http://simbad.cds.unistra.fr/simbad/sim-id?Ident=10+Lac&NbIdent=1&Radius=2&Radius.unit=arcmin&submit=submit+id}{10~Lac} will be observed in an intra-pixel pattern. With this pattern the goal is to derive an empirical correction for the non-linearity and debiasing-induced BFE, applicable to point sources, by deriving a ramp model for each part of the PSF. The ramp model can then be interpolated to calibrate the ramps of any point source science target, this way an optimal flux calibration can be achieved \citep{gasman2022}. This method is already being used to improve the phase curve of HD189733b with the LRS.

Finally we mention that, although the wide use of the developed MIRI electrostatic model for real-time calibration purposes comes with its limitations, such a model would have been extremely useful to have right before and during the ground testing of MIRI. Having a non-linearity and BFE simulator simulating ramps for different illumination profiles allows to optimize calibration algorithms and solutions. It also allows to assess the impact of the effect on the PSF modeling, the absolute flux calibration, as well as on specific science cases of interest. Given the prevalence of the BFE in CCDs, HXRG detectors and Si:As IBC devices, future instrument simulators would benefit greatly from including a electrostatic simulation component.

\section{Conclusions}
\label{sec:conclusions}

A debiasing-induced BFE makes the \textit{JWST} MIRI data yield a 10~--~25~\% larger PSF and LSF as a function of the output level dynamic range covered by the detector pixels during an integration. This affects aperture-corrected photometry, PSF-weighted photometry, PSF subtraction, high-contrast imaging, as well as estimates of kinematics based on spectral line profiles. It also directly affects the calibration of MIRI raw ramp data, translating into a systematic uncertainty on the absolute flux calibration as a function of the exposure parameters used to observe flux calibration standards. 

Using the public \texttt{Poisson\_CCD} code \citep{Lage2021} and a model for the MIRI Si:As IBC sensor infrared-active layer, we found strong evidence for the physical mechanism of the BFE impacting the MIRI pixel voltage integration ramps. We find that this mechanism is driven by detector debiasing due to photoelectron accumulation. This debiasing changes the electrostatic potential inside the infrared-active layer. The paths followed by the excited photoelectrons to the pixels are dictated by the corresponding electric field. At higher levels of debiasing the photoelectrons generated above a given pixel are guided to the neighboring pixels instead. This results in an increase in the slope of the pixel integration ramps as opposed to the decrease in slope recorded for spatially uniform sources. 

While the electrostatic potential results in the electrons being guided to neighboring pixels, the debiasing also results in the width of the depletion region in the infrared-active layer shrinking towards the output (front) contact. This means that more electrons are lost due to recombination, and this constitutes the known MIRI non-linearity effect. We argue that for point sources, and for sources with a significant spectral and spectral flux contrast, the non-linearity and the BFE cannot be considered to be independent. An accurate pixel integration ramp model needs to account for both. Only applying a non-linearity solution derived based on an extended source illumination to non-uniform illumination will result in systematically larger uncertainties in flux calibration.

Having presented a numerical model that predicts the performance of the MIRI Si:As IBC detectors under non-uniform illumination, this forms a basis to understand the ground test and flight MIRI data. MIRI data in flight can be used to refine the presented model, however, the power of the model lies in better interpreting the MIRI flux systematics, deriving an empirical correction based on ramps of reference stars, and informing the detector operation of future infrared missions.

\begin{acknowledgements}
   Ioannis Argyriou and Danny Gasman thank the European Space Agency (ESA) and the Belgian Federal Science Policy Office (BELSPO) for their support in the framework of the PRODEX Programme. Ioannis Argyriou would also like to personally thank Eddie Bergeron (STScI) for his investigations of the Brighter-Fatter Effect across the multiple \textit{JWST} instruments and detectors. \\
   Craig Lage gratefully acknowledges financial support from DOE grant DE-SC0009999 and NSF/AURA/LSST grant N56981CC.\\
   J. A-M acknowledges support by grant PIB2021-127718NB-100 by the Spanish Ministry of Science and Innovation/State Agency of Research MCIN/AEI/10.13039/501100011033 and by “ERDF A way of making Europe”.\\
   Alvaro Labiano acknowledges the support from Comunidad de Madrid through the Atracción de Talento Investigador Grant 2017-T1/TIC-5213, and PID2019-106280GB-I00 (MCIU/AEI/FEDER,UE).\\ 
   The work presented is the effort of the entire MIRI team and the enthusiasm within the MIRI partnership is a significant factor in its success. MIRI draws on the scientific and technical expertise of the following organisations: Ames Research Center, USA; Airbus Defence and Space, UK; CEA-Irfu, Saclay, France; Centre Spatial de Liége, Belgium; Consejo Superior de Investigaciones Científicas, Spain; Carl Zeiss Optronics, Germany; Chalmers University of Technology, Sweden; Danish Space Research Institute, Denmark; Dublin Institute for Advanced Studies, Ireland; European Space Agency, Netherlands; ETCA, Belgium; ETH Zurich, Switzerland; Goddard Space Flight Center, USA; Institute d'Astrophysique Spatiale, France; Instituto Nacional de Técnica Aeroespacial, Spain; Institute for Astronomy, Edinburgh, UK; Jet Propulsion Laboratory, USA; Laboratoire d'Astrophysique de Marseille (LAM), France; Leiden University, Netherlands; Lockheed Advanced Technology Center (USA); NOVA Opt-IR group at Dwingeloo, Netherlands; Northrop Grumman, USA; Max-Planck Institut f\"{u}r Astronomie (MPIA), Heidelberg, Germany; Laboratoire d’Etudes Spatiales et d'Instrumentation en Astrophysique (LESIA), France; Paul Scherrer Institut, Switzerland; Raytheon Vision Systems, USA; RUAG Aerospace, Switzerland; Rutherford Appleton Laboratory (RAL Space), UK; Space Telescope Science Institute, USA; Toegepast- Natuurwetenschappelijk Onderzoek (TNO-TPD), Netherlands; UK Astronomy Technology Centre, UK; University College London, UK; University of Amsterdam, Netherlands; University of Arizona, USA; University of Bern, Switzerland; University of Cardiff, UK; University of Cologne, Germany; University of Ghent; University of Groningen, Netherlands; University of Leicester, UK; University of Leuven, Belgium; University of Stockholm, Sweden; Utah State University, USA. A portion of this work was carried out at the Jet Propulsion Laboratory, California Institute of Technology, under a contract with the National Aeronautics and Space Administration.

   We would like to thank the following National and International Funding Agencies for their support of the MIRI development: NASA; ESA; Belgian Science Policy Office; Centre Nationale D'Etudes Spatiales (CNES); Danish National Space Centre; Deutsches Zentrum fur Luft-und Raumfahrt (DLR); Enterprise Ireland; Ministerio De Economiá y Competividad; Netherlands Research School for Astronomy (NOVA); Netherlands Organisation for Scientific Research (NWO); Science and Technology Facilities Council; Swiss Space Office; Swedish National Space Board; UK Space Agency.

   We take this opportunity to thank the ESA \textit{JWST} Project team and the NASA Goddard ISIM team for their capable technical support in the development of MIRI, its delivery and successful integration.

\end{acknowledgements}

\bibliographystyle{aa} 
\bibliography{aanda} 

\end{document}